\documentclass[useAMS,usenatbib]{mn2e}
\def\lsim{\lower.5ex\hbox{$\; \buildrel < \over \sim \;$}}
\def\gsim{\lower.5ex\hbox{$\; \buildrel > \over \sim \;$}}
\def\be{\begin{equation}}
\def\ee{\end{equation}}

\def\bc{\begin{center}}
\def\ec{\end{center}}

\def\etal{{\em et al.}}
\def\ie{{\em i.e.,}}

\def\etal{{\em et al.}}
\def\ie{{\em i.e.,}}
\def\lsim{\lower.5ex\hbox{$\; \buildrel < \over \sim \;$}}
\def\gsim{\lower.5ex\hbox{$\; \buildrel > \over \sim \;$}}
\def \simeq{\lower.3ex\hbox{$\; \buildrel \sim \over - \;$}}

\def\apj{{ApJ}}%
\def\apjl{{ApJ}}%
\def\aap{{A\&A}}%
\def\mnras{{MNRAS}}%
\def\na{{New A}}%
\def\pasj{{PASJ}}%

\usepackage{epsfig}
\usepackage{graphicx}	
\usepackage{amsmath}	
\usepackage{amssymb}	
\usepackage{color}
\title[Magnetically supported accretion flow around black holes]{Properties of
magnetically supported dissipative accretion flow around black holes
with cooling effects}

\author[Sarkar, B., Das, S., Mandal, S.]{Biplob Sarkar$^{1}$\thanks{E-mail: biplob@iitg.ernet.in (BS);
sbdas@iitg.ernet.in (SD); samir@iist.ac.in (SM)},
Santabrata Das$^1$, Samir Mandal$^2$
\\
$^{1}$Indian Institute of Technology Guwahati (IITG),
 Guwahati 781039, Assam, India\\
$^{2}$Indian Institute of Space Science $\&$ Technology (IIST), Trivandrum 695547, India.
}

\date{Accepted XXX. Received YYY; in original form ZZZ}

\pubyear{2016}

\begin{document}
\date{}
\maketitle
\label{firstpage}

\begin{abstract}
We investigate the global structure of the advection dominated accretion flow
around a Schwarzschild black hole where the accretion disc is threaded by
toroidal magnetic fields. We consider synchrotron radiative process as an
effective cooling mechanism active in the flow. With this, we obtain the global
transonic accretion solutions by exploring the variety of boundary conditions
and dissipation parameters, namely accretion rate (${\dot m}$) and viscosity
($\alpha_B$). The fact that depending on the initial parameters, steady state
accretion flows can possess centrifugally supported shock waves. These global
shock solutions exist even when the level of dissipation is relatively high. We
study the properties of shock waves and observe that the dynamics of the
post-shock corona (hereafter, PSC) is regulated by the flow parameters.
Interestingly, we find that shock solution disappears completely when the
dissipation parameters exceed their critical values. We calculate the critical
values of viscosity parameter ($\alpha^{\rm cri}_B$) adopting the canonical
values of adiabatic indices as $\gamma=4/3$ (ultra-relativistic) and $1.5$
(semi-non-relativistic) and find that in the gas pressure dominated domain, 
$\alpha^{\rm cri}_B \sim 0.4$ for $\gamma=4/3$ and $\alpha^{\rm cri}_B \sim 0.27$
for $\gamma=1.5$, respectively. We further show that global shock solutions
are relatively more luminous compared to the shock free solutions. Also, we
have calculated the synchrotron spectra for shocked solutions. When the shock
is considered to be dissipative in nature, it would have an important implication
as the available energy at PSC can be utilized to power the outflowing matter
escaped from PSC. Towards this, we calculate the maximum shock luminosity
and discuss the observational implication of our present formalism.

\end{abstract}




\section{Introduction}

The presence of magnetic field is ubiquitous in all astrophysical objects
and hence, the accretion discs around black holes are also presumably be
magnetized in nature. In the course of accretion process, the magnetic
field inside the disc is inherited either from the companion star or
from the interstellar medium \citep{Bisnovatyi-Kogan-Ruzmaikin74}
and these fields are considered as `frozen'  with the accreting matter.
In addition, magnetic fields within the disc could originate by the
dynamo action as has been proposed by various authors \citep{torkel94,
Khanna-Camenzind96,Balbus-Hawley98}. Meanwhile, the importance of
magnetic fields in accretion theory was realized by \citet{bl69,ss73}
where they reported out that magnetic fields can contribute to the
angular momentum transport and explain the unknown nature of disc
viscosity. Moreover, \citet{Bisnovatyi-Kogan-Blinnikov76} pointed
out that the presence of magnetic field in the accretion disc seems
to be necessary in order to account for the X-ray and gamma ray
emissions from Cyg-X1.

Numerous studies have now conclusively established that magnetic fields indeed play a very 
important role in the accretion process around black holes. Magnetic fields affect the dynamics 
of an accretion disc in a number of ways. Firstly, the turbulent magnetic fields generates 
Maxwell stress which causes efficient angular momentum transport in the disc leading to faster 
infall of accreting matter into the black hole \citep{Balbus-Hawley91}. Secondly, observational features 
such as bipolar outflows and highly-collimated jets found in a large number of black hole 
candidates can be satisfactorily explained due to the presence of a magnetic field. For example, 
\citet{m06}, reported the evidence of magnetic fields helping to drive a highly ionized wind 
in the stellar mass system GRO J1655-40. Thirdly, heating of the accretion disc may be caused 
by dissipation of magnetic energy via magnetic reconnection \citep{machida06,hirose2006,krolik2007}.
Last but not the least, the presence of magnetic fields lead to efficient cooling of the disc because of 
synchrotron emission by the ionized electrons.

In an accretion disc, differential rotation is the dominant motion over radial infall and this
causes the toroidal component of magnetic field to play a major role. We thus consider an 
accretion disc with turbulent magnetic fields where the azimuthal component of the magnetic 
field is dominant. The characteristic behaviour of the magnetic and gas pressure are considered 
to be the same and the total pressure (i.e. gas plus magnetic) provides the vertical support of 
the disc against gravity. Thus in the standard $\alpha$-prescription of viscosity, if the total 
pressure is substituted, there shall be an enhancement of angular momentum transport in the 
disc since magnetic pressure contributes to the total pressure. Accretion flows with toroidal 
magnetic fields have been studied both analytically as well as through numerically simulations. 
Analytical studies on the effect of toroidal magnetic field on the disc have been carried out 
by \citet{Akizuki-Fukue06,Begelman-Pringle07,bu09,oda07,oda10,oda12,mosall14,sam14,sarkar2015,
Sarkar-Das16}. Numerical 
simulations of accretion disc with toroidal magnetic field around black holes have been carried 
out by \citet{machida06,hirose2006,johansen2008}. \citet{machida06} showed the existence of a 
magnetically supported, quasi-steady disc during the transition from a low/hard state to 
a high/soft state in black hole accretion flows. Recent simulation results by 
\citet{sadowski16} also validate the conclusions of \citet{machida06}.

In order to preserve the inner boundary conditions, flow around a black hole must necessarily 
be transonic. Accreting matter around a black hole feels a centrifugal force because of its rotation. 
Matter flowing supersonically near the black hole slows down due to the 
barrier originating form the centrifugal force. This gives rise to discontinuities in the flow 
variables in the form of shock waves where the flow undergoes a `supersonic to sub-sonic' transition. 
This phenomena of shock formation produces centrifugal pressure supported  virtual barrier 
\citep{c89} that  equivalently refers to the post-shock region (PSC). In absence of the physical 
boundary layer around the black holes, PSC behaves like a effective boundary layer for the accretion flow 
for all practical purposes \citep{Chakrabarti96,dcnc01,Chakrabarti-Das04,d07,cc11}. 
The kinetic energy of the flow suffers a sudden reduction due to the shock transition and the
thermal energy rises. Hence matter in PSC is heated up and also it gets compressed. Since PSC is hot 
and compressed compared to the pre-shock matter, it develops an excess thermal gradient force which 
drives a part of the accreting matter as bidirectional jets/outflows \citep{c99,dcnc01,cd07,Singh-Chakrabarti11,
kc13,Das-etal14,dcns14,Aktar-etal15}. The soft photons from the outer disc interact with the hot 
electrons in the PSC and are inverse Comptonized to higher energies producing observable hard photons 
\citep{ct95,MandalChakrabarti05,ChakrabartiMandal06}. Also modulation of the PSC cause the
phenomena of quasi-periodic oscillations of hard 
radiations in the X-ray spectrum \citep{Chakrabarti-Manickam00}.

Recently, \citet{sarkar2015} have studied magnetized accretion flow around black holes considering 
the electrons to be cooled by synchrotron emission. For a particular set of injection parameters 
and different cooling rates, the flow was shown to undergo a shock transition before entering into 
the black hole. They concluded that apart from viscosity and accretion rate, the plasma $\beta$ 
parameter also has a significant effect on the shock dynamics. \citet{Sarkar-Das16} have studied 
optically thin magnetized accretion flow around stationary black holes assuming Comptonization of 
bremsstrahlung radiation to be the active cooling mechanism. They have established that apart from 
global accretion solutions passing through the inner sonic point studied by \citet{oda07,oda12}, 
solutions which pass through the outer sonic point are also possible and such solutions exist over 
a wide range of outer boundary conditions. In principle, the flow may possess a maximum of three 
sonic points outside the event horizon \citep{c89,dcc01,Chakrabarti-Das04,sarkar2013}. Depending on the outer 
boundary conditions, a flow may either pass through the inner sonic point only or may first pass 
through the outer sonic point and after a shock transition pass through the inner sonic point. Thus, 
the presence of multiple sonic points is a necessary condition for the formation of 
shock in the flow. The authors showed that the energy dissipation at the stationary shocks can 
explain the core radio luminosity of AGNs. Also, the maximum disc luminosity was estimated to be 
exceeding 10\% of the Eddington luminosity. In global solutions, however, the authors considered 
the magnetic field strength to be moderate throughout the flow. The bremsstrahlung cooling mechanism 
employed by \citet{Sarkar-Das16} is the dominant cooling process in the accretion disc around a supermassive 
black hole where the magnetic field is expected to be low because of large size of the accretion disc. 
However, the magnetic field in the accretion disc around a stellar mass black hole is significantly high. 
Since the disc is arrested by significant magnetic field, the hot electrons in the accretion flow around 
stellar mass black holes cool primarily by synchrotron emission. When there is significant cooling of the 
disc, the disc can undergo transition from gas pressure dominated state to magnetic pressure dominated state 
because of flux conservation at a particular radii \citep{oda07,oda12}. The work by \citet{sarkar2015} 
is a preliminary investigation of such magnetized discs around stationary black holes employing the 
synchrotron cooling mechanism. In the present work, motivated by these reasons, we make a detailed 
exploration of the effect of toroidal magnetic fields in magnetically supported accretion discs around 
black holes assuming synchrotron radiation to be the active cooling mechanism. Thus, to model the 
dissipative accretion flow, a set of steady state magneto-hydrodynamical equations are considered. The 
magnetic energy dissipation process governs the heating of the flow. The geometry of space time around 
the black hole is described by the \citet{pw80} pseudo-Newtonian potential. With this, we self-consistently 
calculate the global accretion solution including shock waves and investigate the shock properties in 
terms of the flow parameters. We establish that a wide range of flow parameters admits shocks in an 
accretion flow. We identify the critical limits of viscosity parameter which allows the existence of shocks 
in the entire regime ranging from magnetic pressure dominated to gas pressure dominated disc. We also quantify 
the critical limits of accretion rate which allows the existence of shocks in accretion flow and study its 
variation with plasma $\beta$. Furthermore, we present the overall variation of the typical spectra
for shocked accretion flows injected with a fixed outer edge conditions having different plasma
$\beta$ values. Finally, considering the dissipative nature of shocks, we present the theoretical
estimate of the maximum energy dissipation ($\Delta \mathcal{E}^{\rm max}$) at shock. Using
this available energy at shock, we calculate the maximum kinetic power lost from the disc
($L^{\rm max}_{\rm shock}$), which would be useful to understand the core radio luminosities
associated with black hole sources.

The outline of the paper is as follows. In Section 2, we present the assumptions and governing
equations for our model. In Section 3, we investigate the global accretion solutions with and
without shock, shock properties, and the critical limits of viscosity and accretion rate. Next,
we present the radiation spectra of the shocked accretion flows from our model. We also apply our
formalism to calculate the maximum energy dissipation at shock and the corresponding
shock luminosities. Finally in Section 4, we present the concluding remarks.

\section{Accretion flow model}

We account for the magnetic field structure in the accretion disc based on the results of
numerical simulations of global MHD accretion flow around black holes in the quasi-steady
state \citep{machida06,hirose2006}. These simulations show that magnetic fields inside the
disc are turbulent and dominated by azimuthal component. Hence, based on the findings of
these simulations, the magnetic fields are considered as a combination of mean fields and
the fluctuating fields. We express the mean fields as ${\bf{B}} = (0,<B_{\phi}>,0)$, while the
fluctuating fields are represented by $\delta {\bf{B}} = (\delta B_r,\delta B_{\phi},\delta B_z)$.
Here, $<>$ signifies the azimuthal average. When the fluctuating components are  azimuthally
averaged, we assume that they eventually vanish. Accordingly, the azimuthal component of
magnetic fields dominates over the radial and vertical components as they are negligible,
$|<B_{\phi}> + \delta B_{\phi}|\gg|\delta B_r|$ and $|\delta B_z|$. Basically, this renders the 
azimuthally averaged magnetic field as $<{\bf{B}}>=<B_{\phi}>\hat{\phi}$\citep{oda07}.

\subsection{Governing Equations}
In our current model, we consider a thin, axis-symmetric disc around a Schwarzschild black hole of mass
$M_{BH}$, in the steady state. We use the Geometric unit system as $2G = M_{\rm BH} = c =1$, where, $G$ 
is the universal Gravitational constant and $c$ is the speed of light. In this unit system, length, time 
and velocity are expressed in unit of $r_g = {2GM_{\rm BH}}/{c^2}$, $r_g/c$ and $c$, respectively. 
Further, we adopt cylindrical polar coordinates system ($x,\phi,z$) to represent the accretion 
disc structure where the black hole is located at its origin.

The governing equations that describe the accretion flow around black hole in the steady state are given by:

(a) Radial momentum equation:
$$
\upsilon\frac{d\upsilon}{dx} + \frac{1}{\rho}\frac{dP}{dx} - \frac{\lambda^2(x)}{x^3} +
\frac{d\Psi}{dx} + \frac{\left<B_{\phi} ^2\right>}{4\pi x \rho} = 0.
\eqno(1)
$$
where, $\upsilon$ is the radial velocity, $\rho$ is the density and $\lambda$ is the specific
angular momentum of the flow, respectively. Here, $P$ denotes the total pressure of the 
accretion flow which we consider as $P=p_{gas} + p_{mag}$ where, $p_{gas}$ is the gas
pressure and $p_{mag}$ is the magnetic pressure of the flow, respectively.
The gas pressure inside the disc is given by, $p_{gas} = R \rho T/\mu$, where, $R$ is the
gas constant, $T$ is the temperature and $\mu$ is the mean molecular weight assumed
to be $0.5$ for fully ionized hydrogen. The azimuthally averaged magnetic pressure is
given by $p_{mag} = <B_{\phi}^2>/8\pi$. We define $\beta = p_{gas}/p_{mag}$
and employing this we obtain the total pressure as $P = p_{gas} (\beta + 1)/\beta$.
Here, $\Psi$ represents the potential energy. To mimic the general relativistic effects
around a stationary black hole, we adopt the pseudo-Newtonian potential (\cite{pw80})
which is given by,
$$
\Psi = -\frac{1}{2(x -1)}.
\eqno(2)
$$
This potential satisfactorily describes the dynamical aspects of general relativity effects
in the range $x>1$ and immensely simplifies the basic equations describing the flow
motion in an accretion disc. The last term on the left hand side represents the magnetic
tension force.

(b) Mass Conservation:
$$
\dot{M}=2\pi x\Sigma \upsilon,
\eqno(3)
$$
where, $\dot{M}$ represents the mass accretion rate which is a global constant along the flow. $\Sigma$ 
denotes the vertically integrated density of flow \citep{mkfo84}.

(c) Azimuthal momentum equation:
$$
\upsilon\frac{d\lambda(x)}{dx}+\frac{1}{\Sigma x}\frac{d}{dx}(x^2T_{x\phi}) = 0,
\eqno(4)
$$
where, the vertically integrated total stress is assumed to be dominated by the $x\phi$
component of the Maxwell stress $T_{x\phi}$. Following the work of \citet{machida06},
we estimate $T_{x\phi}$ for an  an advective flow with significant radial velocity \citep{Chakrabarti-Das04} as

$$
T_{x\phi} = {\frac{<B_{x}B_{\phi}>}{4\pi}}h = -\alpha _{B}(W + \Sigma \upsilon^2),
\eqno(5)
$$
where, $h$ denotes the the half thickness of the disc.
In the right hand side of Eq. (5), $\alpha_B$ is the constant of proportionality and $W$ is
the vertically integrated pressure \citep{mkfo84}, respectively. In the current model, we
take $\alpha_B$ as a global parameter throughout the flow as is considered in \citet{ss73}.
When radial advection is negligible, such as for a Keplerian flow, Eq. (5) simplifies to the
original prescription of `$\alpha-$model' \citet{ss73}.

Considering the flow to be in
hydrostatic equilibrium in the vertical direction, $h$ is expressed as,

$$
h = \sqrt{\frac{2}{\gamma}} a x^{1/2} (x - 1),
\eqno(6)
$$
where, $a$ denotes the adiabatic sound speed which is defined as $a=\sqrt {\gamma P/\rho}$,
where $\gamma$ is the adiabatic index. We treat $\gamma$ as a global parameter and adopt
its the canonical value as $\gamma=4/3$ in the subsequent analysis, until otherwise stated.

(d) The entropy generation equation:

$$
\Sigma \upsilon T \frac {ds}{dx}=\frac{h\upsilon}{\gamma-1}
\left(\frac{dp_{gas}}{dx} -\frac{\gamma p_{gas}}{\rho}\frac{d\rho}{dx}\right)=Q^- - Q^+,
\eqno(7)
$$
where, $s$ and $T$ denote the specific entropy and the local temperature of the flow, respectively. 
Here, $Q^+$ and $Q^-$ represent the vertically integrated heating and cooling rates. The heating of the
flow takes place due to the thermalization of magnetic energy through the magnetic reconnection
mechanism \citep{hirose2006,machida06} and therefore, is given by,

$$
Q^{+} = \frac{<B_{x}B_{\phi}>}{4\pi} xh \frac{d\Omega}{dx} = -\alpha _{B}(W + \Sigma \upsilon^2) x \frac{d\Omega}{dx},
\eqno(8)
$$
where, $\Omega$  refers to the local angular velocity of the flow.

In an accretion disc, the cooling of the flow is governed by the various radiative processes,
such as, bremsstrahlung, synchrotron or Comptonization of bremsstrahlung and
synchrotron photons. In this work, since the accretion flow is magnetized in nature,
synchrotron process becomes effective to cool the flow and the cooling rate due to 
the synchrotron radiation is given by  \citep{shte83},

$$
Q^-= \frac{Sa^5\rho h}{\upsilon x^{3/2}(x-1)} \frac{\beta^2}{(1+\beta)^3},
\eqno(9)
$$
with,

$$
S= 1.048 \times 10^{18} \frac{ {\dot m} \mu^2 e^4}{I_n m_e^3\gamma^{5/2}}
\frac{1}{2GM_{\odot}c^3},
$$
where,  $e$ and $m_e$ denote the charge and mass of the electron, respectively. And,
$k_B$ is the Boltzmann constant,
$\mu$ is the mean molecular weight, $I_n = (2^n n!)^2/(2n + 1)!$ and $n = 1/(\gamma - 1)$.
Following \citet{cc02}, we estimate the electron temperature  as $T_e = \sqrt{m_e/m_p}T_p$
where, the coupling between the ions and electrons, if any, is ignored and $m_p$ refers to
the mass of the ion. In the subsequent sections,
we express accretion rate in  units of Eddington rate
(${\dot M}_{\rm edd} = 1.39 \times10^{17} \times M_{BH}/M_{\odot}~{\rm g~s}^{-1}$)
and we denote it by $\dot m$.
Furthermore, in account of the energy loss of the flow, we neglect the contribution
of the bremsstrahlung emission because it is regarded as a very inefficient cooling process
for stellar mass black hole systems \citep{cc00,d07}.

(e) Radial advection of the toroidal magnetic flux:

We describe the advection rate of the toroidal magnetic flux by considering the induction
equation and it is expressed as,

$$
\frac {\partial <B_{\phi}>\hat{\phi}}{\partial t} = {\bf \nabla} \times
\left({\vec{\upsilon}} \times <B_{\phi}>\hat{\phi} -{\frac{4\pi}{c}}\eta {\vec{j}}\right),
\eqno(10)
$$
where, $\vec{\upsilon}$ denotes velocity vector, $\eta$ is the resistivity and
${\vec{j}} = c({\nabla} \times <B_{\phi}>\hat{\phi})/4\pi$ refers the current density.
Here, Eq. (10) is azimuthally averaged. 
Since the Reynolds number ($R_m$) is very large in the case of accretion disc, on
account of very large length scale, we neglect the magnetic-diffusion term.
Moreover, we neglect the dynamo term in the present study. In the steady state,
the resulting equation is then vertically averaged assuming that the averaged
toroidal magnetic fields vanish at the disc surface. With this, we obtain the advection
rate of the toroidal magnetic flux as \citep{oda07},

$$
\dot{\Phi} = - \sqrt{4\pi}\upsilon h {B}_{0} (x) ,
\eqno(11)
$$
where,
\begin{eqnarray*}
{B}_{0} (x) && = \langle {B}_{\phi} \rangle \left(x; z = 0\right)  \nonumber \\
&& = 2^{5/4}{\pi}^{1/4}(R T/\mu)^{1/2}{\Sigma}^{1/2}h^{-1/2}{\beta}^{-1/2}
\end{eqnarray*}
represents the azimuthally averaged toroidal magnetic field confined in the disc
equatorial plane. In eq. (11), the magnetic flux advection rate ($\dot{\Phi}$) is
expected to vary in the radial direction when both the dynamo term and
the magnetic diffusion term is present. Global three-dimensional MHD simulation
of \citet{machida06} indicates that  the magnetic flux advection rate varies as
$\dot{\Phi} \propto 1/x$, when the disc is in quasi steady state. In the present
context, the computation of the magnetic diffusion term and the dynamo terms
is beyond the scope the paper and therefore, based on the above findings, 
we adopt a parametric relation between $\dot{\Phi}$ and $x$ which is given
by \citep{oda07},

$$
\dot{\Phi}\left(x; \zeta, \dot{M}\right) \equiv \dot{\Phi}_{edge}(\dot{M})\left(\frac{x}{x_{edge}} \right)^{-\zeta},
\eqno(12)
$$
where, $\dot{\Phi}_{edge}$ is  the advection rate of the toroidal magnetic flux at the outer edge 
of the disc ($x_{edge}$). When $\zeta = 0$, the conservation of  magnetic flux in the radial 
direction is restored. For $\zeta > 0$, the magnetic flux increases as the accreting matter
approaches the black  hole horizon. In this analysis, we consider $\zeta$ as a global constant
and fix its value as $\zeta = 1$ for representation, unless stated otherwise.

\subsection{Sonic Point Analysis}

In the process of accretion on to the black hole, infalling matter starts its journey from
the outer edge of the disc with negligible radial velocity and subsequently crosses the
black hole horizon with the velocity comparable to the speed of light. 
This findings evidently demand that during accretion, infalling matter must change its sonic
character smoothly
from subsonic state to supersonic state before falling in to the black hole. The radial coordinate
where accreting matter encounter such sonic transition is known as sonic point. In the next,
we carry out the sonic point analysis of the accretion flow by simultaneously solving
equations (1), (3), (4), (7), (11) and (12) which is expressed as,

$$
\frac {d\upsilon}{dx}=\frac{N}{D},
\eqno(13)
$$
where, the numerator $N$ is given by,

$$
N = \frac {Sa^5}{\upsilon x^{3/2}(x-1)}\frac{\beta^{2}}{(1 + \beta)^{3}} +\frac {2\alpha^2_B I_n  (a^2g+\gamma \upsilon^2)^2}{\gamma^2 x \upsilon}
$$
$$
+\frac {2\alpha^2_B g I_n a^2(5x-3)(a^2g+\gamma \upsilon^2)}{\gamma^2 \upsilon x(x-1)}
$$
$$
-\left[ \frac {\lambda^2}{x^3}-\frac {1}{2(x-1)^2}\right]
\left[\frac {[3+\beta(\gamma+1)]\upsilon}{(\gamma-1)(1+\beta)}
 -\frac{4\alpha^2_B g I_n (a^2g+\gamma \upsilon^2)}{\gamma \upsilon} \right]
 $$
 $$
- \frac {\upsilon a^2(5x-3)}{x(\gamma-1)(x-1)}\frac{(\beta + \frac{3}{2\gamma})}{(1 + \beta)} -\frac {4\lambda \alpha_B I_n (a^2g+\gamma \upsilon^2)}{\gamma x^2}
$$
$$
-\frac {8 \alpha_B^2 I_n a^2 g(a^2g+\gamma \upsilon^2)}{\gamma^2 \upsilon (1 + \beta) x} +\frac {2[3 + \beta(\gamma + 1)]a^2 \upsilon}{\gamma(\gamma - 1)(1 + \beta)^2 x}
$$
$$
-\frac {a^2 \upsilon}{\gamma(\gamma - 1)(1 + \beta)(x-1)} - \frac {a^2 \upsilon (4\zeta - 1)}{2\gamma x(\gamma - 1)(1 + \beta)}
\eqno(13a)
$$

and the denominator $D$ is given by,

$$
D = \frac {2a^2}{(\gamma-1)}\left[ \frac{2}{\gamma(1+\beta)}+\frac {\beta}{1+\beta}\right]-\frac {[3+(\gamma+1)\beta]\upsilon^2}
{(\gamma-1)(1+\beta)}
$$
$$
+\frac{2\alpha^2_B I_n (a^2g+\gamma \upsilon^2)}{\gamma}
\left[ (2g-1)-\frac {a^2g}{\gamma \upsilon^2}\right].
\eqno(13b)
$$
Here, we write $g = I_{n+1}/I_n.$

Further, we obtain the gradient of sound speed, angular momentum and plasma $\beta$,
respectively as:

$$
\frac{da}{dx} = \left( \frac{a}{\upsilon} - \frac{\gamma \upsilon}{a} \right)
\frac{d\upsilon}{dx} + \frac{\gamma}{a}\left[ \frac {\lambda^2}{x^3}-\frac {1}{2(x-1)^2}\right]
$$
$$ 
+\frac{(5x-3)a}{2x(x-1)} - \frac{2a}{(1+\beta)x},
\eqno(14)
$$

$$
\frac{d\lambda}{dx} =-\frac{\alpha_{B} x (a^2g- \gamma \upsilon^2)}{\gamma \upsilon^2}\frac{d\upsilon}{dx} +\frac{2 \alpha_{B} axg }{\gamma \upsilon}\frac{da}{dx}  \nonumber \\
$$
$$
+\frac{\alpha_{B}(a^2g+\gamma \upsilon^2)}{\gamma \upsilon},
\eqno(15)
$$

$$
\frac{d\beta}{dx} =\frac{(1+\beta)}{\upsilon}\frac{d\upsilon}{dx} + \frac{3 (1+\beta)}{a}\frac{da}{dx}
+\frac{1+\beta}{x-1}
$$
$$
+\frac{(1+\beta)(4\zeta-1)}{2x}.
\eqno(16)
$$

It is already pointed out that the trajectory of the accretion flow around black hole must be
smooth along the streamline and hence, $d\upsilon/dx$ must be real and finite always. 
From equation (13b), however we infer that the denominator $D$ may vanish at some radial
distance between the outer edge of the disc and the horizon. 
In order to maintain
the flow to be smooth everywhere along the streamline, the point where $D$ goes to
zero, $N$ must also tends to zero there. Indeed, the location where both $N$ and $D$
vanish simultaneously is called as sonic point as the infalling matter becomes transonic
there. Following this, we obtain two conditions at the sonic point as $N =0$ and  $D = 0$.
Setting $D$ to zero, we have the expression of Mach number
($M = \upsilon/a$) at the sonic point ($x_c$) as,

$$
M(x_c) =\sqrt {\frac{-m_2 - \sqrt{m^2_2-4m_1 m_3}}{2m_1}},
\eqno(17)
$$
where,
$$ 
m_1=2\alpha^2_{B} I_n \gamma^2(1 + \beta_c)(\gamma-1)(2g-1) - \gamma^2(3 + (\gamma+1)\beta_c)
$$
$$
m_2=2\gamma(2 + \gamma\beta_c) + 4\alpha^2_{B} I_n \gamma g (1 + \beta_c)(g-1)(\gamma-1)
$$ 
$$
m_3=-2\alpha^2_{B} I_n g^2 (1 + \beta_c)(\gamma-1)
$$ 

Using the remaining sonic point condition $N = 0$, we get an algebraic equation
of the sound speed at $x_c$ which is given by,

$$
{\mathcal A}a^3(x_c) + {\mathcal B}a^2(x_c) + {\mathcal C}a(x_c) +{\mathcal D}= 0 ,
\eqno(18)
$$
where,
$$
{\mathcal A}=\frac{S}{x_c^{3/2}(x_c-1)}\frac{\beta_c^{2}}{(1+\beta_c)^{3}},
$$
$$
{\mathcal B} =  \frac {2\alpha^2_B I_n (g+\gamma M_c^2)^2}
{\gamma^2 x_c}+\frac {2\alpha^2_B I_n g (5x_c-3)(g+\gamma M_c^2)}
{\gamma^2 x_c(x_c-1)} 
$$
$$
-\frac{M_c^2(5x_c-3)}{x_c(\gamma-1)(x_c-1)}\frac{(\beta_c + \frac{3}{2\gamma})}
{(1+\beta_c)} -\frac {8\alpha^2_B I_n g(g+\gamma M_c^2)}
{\gamma^2 (1+\beta_c)x_c}
$$
$$
 + \frac{2[3+\beta_c(\gamma+1)]M_c^2}{\gamma(\gamma-1)(1+\beta_c)^2 x_c} - 
 \frac{M_c^2}{\gamma(\gamma-1)(1+\beta_c)(x_c-1)}
$$
$$
 - \frac{(4\zeta - 1)M_c^2}{2\gamma(\gamma-1)(1+\beta_c) x_c},
$$

$$
{\mathcal C} = -\frac {4\lambda_c \alpha_B I_n M_c (g+\gamma M_c^2)}{\gamma x_c^2},
$$

$$
{\mathcal D} = -\left[ \frac {\lambda_c^2}{x_c^3}-\frac {1}{2(x_c-1)^2}\right] 
$$
$$
\times
\left[\frac {[3+\beta_c(\gamma+1)]M_c^2}{(1+\beta_c)(\gamma-1)}-
\frac{4\alpha^2_B g I_n (g+\gamma M_c^2)}{\gamma} \right].
$$
Here, the quantities with  subscript `c' indicate their values measured at the sonic point.

Upon solving equation (18) for a set of input parameters of the flow, we calculate the
sound speed at $x_c$ and then, find the radial velocity of the flow using equation (17).
Using these quantities in equation (13), we investigate the properties of the sonic point.
At $x_c$, $d\upsilon/dx$ generally possesses two values; one corresponds to accretion
flow and the other for wind solution. When both the values of  $d\upsilon/dx$ are real
and of opposite sign, such a sonic point has special significance as the global transonic
solution only pass through it. Sonic point of this kind is called the saddle type sonic point
\citep{Chakrabarti-Das04}. In this work, we are mainly interested to study the properties of the global
accretion flow around the black holes and hence, the wind solutions are left aside 
for future study.

\section{Results}

We simultaneously solve the differential equations (13), (14), (15) and (16) for a given set of
flow parameters to obtain the global transonic accretion solution around black hole.
In this analysis, 
${\dot m}$, $\alpha_{B}$ and $\gamma$ are treated as the global parameters of the flow.
And, three additional parameters are required in order to start the numerical integration of
the above equations from a reference point ($x_{\rm ref}$). These three parameters are 
$x_{\rm ref}$, angular momentum at $x_{\rm ref}$ ($\lambda_{\rm ref}$) and 
plasma $\beta$ at $x_{\rm ref}$ ($\beta_{\rm ref})$, respectively and they are treated as
the local parameters. With this, we obtain the transonic global accretion solutions that
may contain shock waves. It is to be noted that we present 
angular momentum ($\lambda_{\rm ref}$) in terms of Keplerian angular momentum 
$\lambda_K~(\equiv \sqrt{x_{\rm ref}^3/2(x_{\rm ref}-1)^2})$ 
all throughout the paper.

\subsection{GLOBAL ACCRETION SOLUTIONS}

\begin{figure}
\begin{center}
\includegraphics[width=0.45\textwidth]{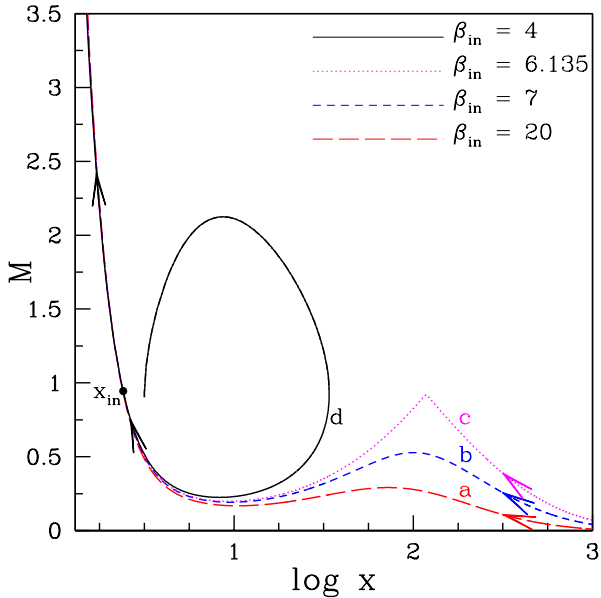}
\end{center}
\caption{Radial dependence of Mach number ($M = \upsilon/a$) of the accreting matter
for different values of plasma $\beta$ ($\beta_{\rm in}$) at the inner sonic point
$x_{\rm in} = 2.41$. Here, $\lambda_{\rm in} = 0.9327\lambda_K$, 
$\alpha_B = 0.02$ and $\dot{m} = 0.004$. 
Long dashed and short dashed curves represent the results for $\beta_{\rm in} = 20$ and $7$, 
respectively. For the same set of flow parameters, the minimum values of plasma $\beta$
that provides the accretion solution connecting the inner sonic point to up to a large distance
(equivalently  `outer edge of the disc') is obtained as $\beta_{\rm in}^{cri} = 6.135$ and the 
solution is depicted with the dotted curve. When $\beta_{\rm in} < \beta_{\rm in}^{cri}$,
accretion solutions fail to connect to the outer edge of the disc and such a representative
solution is depicted  by the solid curve for $\beta_{\rm in} = 4$.}
\end{figure}

In Fig. 1, we present the `phase space diagrams' of accretion
solutions where the Mach number ($M = \upsilon/a$) is plotted as function of the logarithmic
radial distance ($x$). 
Here, we consider the inner sonic point ($x_{\rm in}$) as the reference point ($x_{\rm ref}$) and
choose $x_{\rm in} = 2.41$.
The angular momentum and  the plasma $\beta$ at $x_{\rm in}$ are kept fixed as
$\lambda_{\rm in} = 0.9327\lambda_K$
and $\beta_{\rm in}=20$, respectively. Moreover, we choose 
$\alpha_B = 0.02$, $\dot{m} = 0.004$ and $\gamma=4/3$. With this set of parameters,
we integrate equations (13-16) starting from the inner sonic point once inward up to
black hole horizon and then outward up to a large distance (equivalently  `outer edge
of the disc') . Upon joining them, ultimately we get a global transonic accretion
solution as it connects the outer edge of the disc with the black hole horizon.
In the figure, this solution is plotted with long-dashed curve and marked as `a'. The arrow 
indicates the direction of the flow. Next, we decrease plasma $\beta$ at $x_{\rm in}$ as
$\beta_{\rm in} = 7$ and calculate the global transonic accretion solution keeping the
remaining flow parameters fixed. The obtained result is plotted using short dashed
curve and marked with `b'. We observe that the qualitative feature of the solution
`b' is very much similar to solution `a'.  With the gradual decrease of $\beta_{\rm in}$,
we obtain a critical value $\beta_{\rm in}^{cri} = 6.135$ corresponding to the chosen
set of flow parameters, which is the minimum value of $\beta_{\rm in}$ that provides
a transonic global accretion solution passing through the inner sonic point. This solution
is plotted with the dotted curve and marked as `c'. Needless to mention that the
curves marked with `a-c' represent the results identical to the well known
advection dominated accretion flow (ADAF) solutions around black holes \citep{nar97,oda07}.
When $\beta_{\rm in} < \beta_{\rm in}^{cri}$, transonic accretion solution fails
to extend up to the outer edge of the disc and does not represent a complete
global accretion solution. For representation, we obtain such a solution for $\beta_{\rm in} = 4$
which is illustrated with solid curve and marked as `d'.

Apparently, solution `d' in Fig. 1 does not connect the black hole horizon with the
outer edge of the disc. However, solutions of this kind are potentially promising in
the sense that they can be joined with the another transonic accretion solution
passing through the outer sonic point ($x_{\rm out}$) via shock transition. The
complete description of such a  composite global accretion solution can be visualized
in the following manner. 
The rotating subsonic accretion flow at the outer edge of the disk slowly gains its
radial velocity due to the attraction of gravity. Subsequently, flow becomes
supersonic after passing through $x_{\rm out}$ and continues to proceed towards
the black hole.  In the vicinity of the black holes, since the viscous timescale
greatly exceeds over the infall timescale, the angular momentum transport due
to viscosity becomes feeble and therefore, centrifugal repulsion becomes
comparable to the gravitational force there. As a result, infalling matter
experiences a virtual barrier and eventually slows down.  This causes the piling
of matter and develops local turbulence that establishes entropy generation
leading towards the triggering of shock transition provided the shock conditions
are satisfied.
In fact, according to the second law of thermodynamics, the presence of
the shock wave in accretion flows is thermodynamically preferred when the
post-shock flow possesses high entropy content \citep{Becker-Kazanas01}.
The entropy content of a dissipative accretion flow is obtained as \citep{Chakrabarti96},
$$
\dot{\cal{M}}(x) \propto \left(\frac{\beta}{1 + \beta}\right)^n a^{(2n+1)}\upsilon x^{3/2} (x - 1),
\eqno(19)
$$
where $n = 1/(\gamma - 1)$ is the polytropic index of the flow. In the absence of any dissipative
processes, namely viscosity and/or radiative cooling, $\dot{\cal{M}}$  remains constant throughout
the flow except at the shock transition.

In an accretion disk, transition of flow variables in the form of shock wave is
manifested through the conservation laws of 
mass, momentum, energy and magnetic field \citep[and reference therein]{Sarkar-Das16}. Across the
shock front, these laws are explicitly expressed as the continuity of (a) mass
flux (${\dot M}_{-}={\dot M_{+}}$) (b) the momentum flux ($W_{-}+\Sigma_{-}
\upsilon^2_{-}=W_{+}+\Sigma_{+} \upsilon^2_{+}$) (c) the energy flux 
(${\cal E_{-}}={\cal E_{+}}$) and (d) the magnetic flux ($\dot {\Phi}_{-}=\dot {\Phi}_{+}$),
respectively where, the quantities with subscripts `-' and `+' refer their values
before and after the shock. Here, we calculate the local energy of the flow as
\citep{fukue90,sam14},
$$
\mathcal{E}(x) = \frac{\upsilon^2}{2} + \frac{a^2}{\gamma-1} + 
\frac{\lambda^2}{2x^2} - \frac{1}{2(x-1)} + \frac{<B_\phi ^2>}{4\pi \rho},
\eqno(20)
$$
where, the all quantities have their usual meaning. Employing these shock
conditions, in the next, we calculate the shock location and its various properties
knowing the flow parameters.

%
\begin{figure}
\begin{center}
\includegraphics[width=0.45\textwidth]{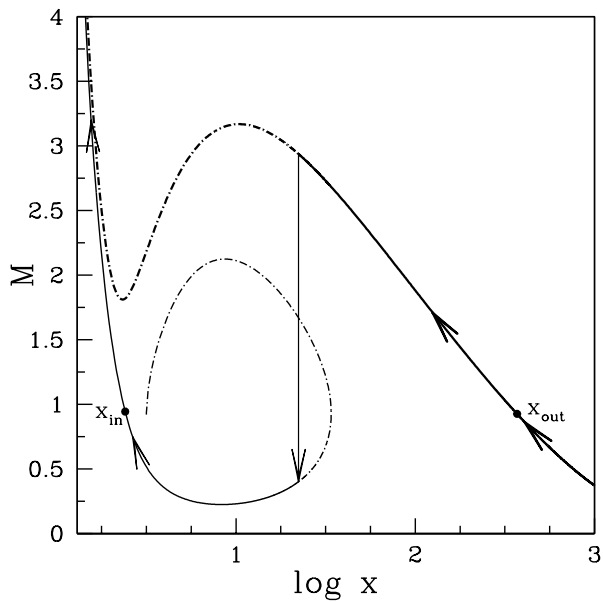}
\end{center}
\caption{
Illustration of a complete global accretion solution including shock transition in the
${\rm log}~ x - M$ plane. Inner sonic point ($x_{\rm in}$) and outer sonic point ($x_{\rm out}$)
are marked. Arrows indicate the direction of the flow motion and vertical arrow represents 
the shock transition. Flow parameters used here are same as Fig. 1d. See text for details.
}
\end{figure}

Toward this, we choose the flow parameters at the inner sonic point ($x_{\rm in}$)
as in the case `d' of Fig. 1. Following the method described in \citet{Chakrabarti-Das04}, 
using the shock conditions, we then uniquely determine the outer sonic point location
($x_{\rm out}$) and other flow variables of the accretion flow at $x_{\rm out}$
for the chosen inner sonic point. 
Utilizing these flow variables at $x_{\rm out}$, we integrate equations (13-16)
outward up to the outer edge of the disc (chosen as $x=1000$) in order to obtain
a complete 
global accretion solution including shock waves. We present the result in Fig. 2,
where Mach number ($M$) is plotted as function of logarithmic radial distance.
In reality, the obtained solution is needed to be visualized in the following way.
The inflowing matter is injected subsonically from the outer edge of the disc
at $x_{\rm edge} = 1000$. Flow becomes supersonic after crossing the outer sonic point at 
$x_{\rm out} = 369.98$ and continue to proceed towards the horizon. This is shown 
by thick solid curve.
Flow may continue its journey even further as shown by thick dotted curve, but eventually 
shock conditions are satisfied at $x_s = 22.36$ and hence, supersonic flow undergoes
a discontinuous transitions in the form of shock waves to become subsonic. This is
indicated by the vertical arrow. Due to gravity, subsonic flow again gains its radial 
velocity and ultimately enters in to the black hole supersonically after passing through
the inner sonic point at $x_{\rm in}=2.41$. This part of the solution is depicted using thin
solid curve. In the figure, $x_{\rm in}$ and $x_{\rm out}$ are marked. Arrows indicate
the overall direction of the flow motion towards the black hole.

%
\begin{figure}
\begin{center}
\includegraphics[width=0.45\textwidth]{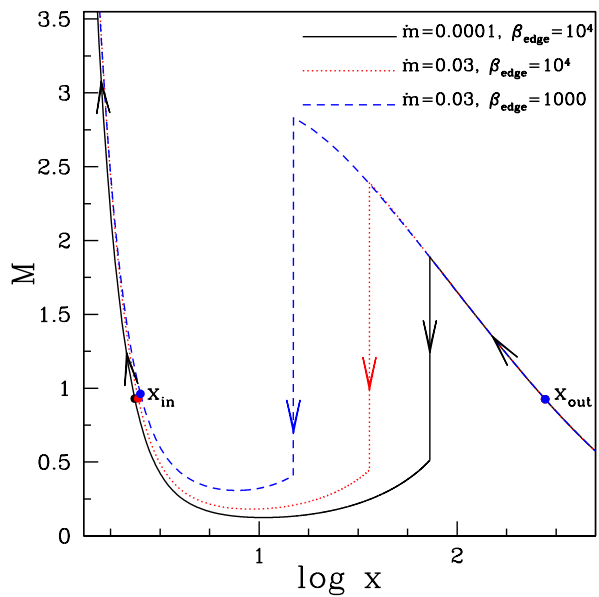}
\end{center}
\caption{
Variation of Mach number as function of logarithmic radial distance. Flows
are injected from the outer edge $x_{\rm edge} = 500$ with energy
$\mathcal{E}_{\rm edge} = 1.025 \times 10^{-3}$, angular momentum 
$\lambda_{\rm edge}= 0.1386\lambda_K$
and viscosity $\alpha_B = 0.02$. Solid,
dotted and dashed curves depict the results obtained for
(${\dot m}$, $\beta_{\rm edge}$) = ($10^{-4}, 10^{4}$), ($0.03, 10^{4}$)
and ($0.03, 10^{3}$), respectively.
Vertical arrows indicate the corresponding shock transitions positioned
at $x_s = 72.79$ (solid), $x_s = 36.11$ (dotted) and $x_s = 14.88$ (dashed).
See text for details.}
\end{figure}

\begin{center}
\begin{table}
\caption{Flow variables measured at the sonic points for a shock induced global accretion 
solution. See text for details.}
\begin{tabular}{l c c c c c}
\hline\hline
Sonic & $x_c$ & $\lambda_c$ & $\beta_c$ & $v_c$ & $a_c$ \\
Point &  &  &  & &  \\\hline
Inner & 2.3550 & 0.9392$\lambda_K$ & 24.9315 & 0.15333 & 0.16495 \\\hline
Outer & 278.3447 & 0.1706$\lambda_K$  & 6132.3040 & 0.02648 & 0.028598 \\
\hline
\end{tabular}\\
Note: Subscript `c' refers to the quantities measured at sonic points. For inner (outer) sonic
point, `c' is identified with `in' (`out').
\end{table}
\end{center}

In the next, we examine the effects of the  dissipation parameters
($\beta_{\rm edge}$ and/or $\dot{m}$) on the dynamics of shock location
for flows with fixed initial parameters. In Fig. 3, inflowing matter is
injected subsonically from the outer edge of the disc at $x_{\rm edge} = 500$
with specific energy $\mathcal{E}_{\rm edge} = 1.025 \times 10^{-3}$,
angular momentum $\lambda_{\rm edge}= 0.1386\lambda_K$ and $\alpha_B = 0.02$.
When the accretion rate and the plasma $\beta$ at $x_{\rm edge} = 500$
are chosen as $\dot{m} = 0.0001$ and $\beta_{\rm edge} = 10^4$ respectively,
flow encounter shock transition at $x_s = 72.79$. In the figure, solid curve
represents this result where the vertical arrow indicates the shock location.
The corresponding inner and outer sonic points flow variables are given in table 1.
Next, we increase the accretion rate as $\dot{m} = 0.03$, keeping all the remaining flow
parameters fixed at $x_{\rm edge}$ and observe that the shock front moves towards the
horizon at $x_s = 36.11$. This result is depicted using dotted curve where the dotted vertical
arrow represents the shock transition. Increase of $\dot{m}$ evidently
enhances the cooling rate of the flow.
Since density and temperature of the flow are boosted up in the post-shock flow (PSC),
effect of cooling at PSC is more intense compared to pre-shock flow and therefore, 
it reduces thermal pressure that 
cause the shock front to move close to the horizon
in order to maintain pressure balance across the shock front. Further, we fix
$\beta_{edge} = 1000$ and $\dot{m} = 0.03$ keeping the other flow parameters unchanged
at $x_{\rm edge}$ and obtain the shock location at $x_s = 14.88$. We plot this result by dashed
curve where dashed vertical line denotes shock location as before.
Lowering of $\beta_{edge}$ demonstrates the increase of  turbulent magnetic fields in the 
accretion  flow that eventually leads to increase of Maxwell stress and therefore, the 
angular momentum transport from the inner part of the disc to the outer part of the disc
is enhanced. This results the weakening of the centrifugal repulsion at PSC. Moreover,
decrease of $\beta_{edge}$ eventually increases the synchrotron cooling efficiency as well.
Overall, with the combined effects of both physical processes, shock front is pushed 
even further towards the horizon. This exhibits the fact that apart from ${\dot m}$, the role of
$\beta_{edge}$ is also important in determining the dynamics of shock location.

\begin{figure}
\begin{center}
\includegraphics[width=0.45\textwidth]{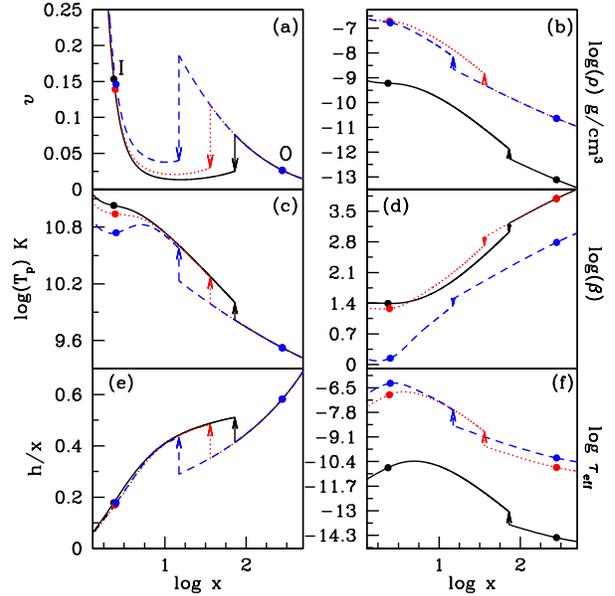}
\end{center}
\caption{
Variation of (a) radial velocity, (b) density in g/cm$^3$, (c) Temperature (d) ratio of gas
pressure to magnetic pressure, (e) disc scale height ($h/x$) and (f) effective optical
depth as function of logarithmic radial coordinate. Results plotted with solid, dotted
and dashed curves correspond to the accretion solution depicted in Fig. 3. Filled circles 
represent the sonic points where the closer one is the inner sonic point and the furthest 
one is the outer sonic point. Vertical arrows indicate the shock position. See text for details.}
\end{figure}

In Fig. 4, we illustrate the vertically averaged accretion disc structure corresponding to
the solutions depicted in Fig. 3. In each panel, we plot the variation of a flow variable with
logarithmic radial distance where the  vertical arrows indicate the shock transition. 
Here, we consider $M_{\rm BH} = 10M_{\odot}$ as a fiducial value.
In Fig. 4a, we demonstrate the radial velocity profile
($\upsilon$) of the accretion flow. As expected, $\upsilon$ increases
with the decrease of radial coordinate until it undergoes a shock transition. Across the shock,
$\upsilon$ drops down to a subsonic value and again increases gradually in the post-shock region. Finally 
flow enters in to the black hole with velocity comparable to the speed of light after passing
through the inner sonic point. Here, solid, dotted and dashed curves represent results
for ($\dot{m}, \beta_{\rm edge}$) $= (10^{-4}, 10^{4})$, $(0.03, 10^{4})$ and $(0.03, 10^{3})$,
respectively. We show the density profile of the flow in Fig. 4b, where we observe the rise
of density immediately after the shock transition in every cases. This happens due to the
the reduction of radial velocity at PSC and eventually preserves the conservation of
mass flux across the shock front.  The overall density profile corresponding to ${\dot m}=0.03$
is higher compared to the case of ${\dot m}=10^{-4}$ simply because the large ${\dot m}$  stands for 
higher mass inflow at the outer edge.
In Fig. 4c, the proton temperature profile ($T_p$) is shown. Across the shock front, 
supersonic pre-shock flow is turned into the subsonic flow and therefore, most of
the kinetic energy of the infalling matter is converted to the thermal energy at PSC.
This eventually leads  to the heating of the PSC as indicated by the rise of post-shock
temperature profile. Interestingly, when ${\dot m}$ is increased,  the effect of cooling
becomes more effective at PSC that causes the reduction of $T_p$ as clearly seen 
close to the inner part of the disc. In addition, we observe that the decrease of
$\beta_{\rm edge}$ essentially exhibits the reduction of the temperature profile 
of the accretion flow. This finding is in agreement with the results of numerical 
simulation \citep{machida06} which yields a cooler disk structure for a magnetically
dominated accretion flow. In Fig. 4d, we show the radial variation of plasma $\beta$.
We find that $\beta$ decreases as the flow approaches to the black hole. Moreover,
$\beta$ falls down sharply across the shock that turns PSC into magnetically dominated.
The radial dependence of the vertical scale-height ($h/x$) is presented in Fig. 4e. The
validity of the thin disc approximation is observed all throughout from the outer
edge to the horizon even in the presence of shock transition. Finally,  in Fig. 4f, we
demonstrate the variation of effective optical depth, 
$\tau_{\rm eff} = \sqrt{\tau_{\rm es}\tau_{\rm syn}}$ where, $\tau_{\rm es}$
represents the scattering optical depth given by $\tau_{\rm es} = \kappa_{\rm es} \rho h$
and the electron scattering opacity, $\kappa_{\rm es}$, is taken to be
$\kappa_{\rm es} = 0.38~{\rm cm}^2 {\rm g}^{-1}$. Here, $\tau_{\rm syn}$ denotes
the absorption effect arising due to thermal processes and is given by
$\tau_{\rm syn} =\left( h q_{\rm syn}/4 \sigma T_{e}^4\right)\left(2GM_{BH}/c^2\right)$
where, $q_{\rm syn}$ is the synchrotron emissivity \citep{shte83} and $\sigma$
is the Stefan-Boltzmann constant. We find that the optical depth of PSC is always
greater than the pre-shock region as the density in the post-shock region is higher
(see, Fig. 4b). In addition, the overall variation of the optical depth for enhanced
accretion rate remains higher all throughout.
Moreover, in spite of the steep
density profile, PSC is found to remain optically thin ($\tau < 1$). This apparently indicates
that the possibility of escaping hard radiations from the PSC seems to be significant.

\begin{figure}
\begin{center}
\includegraphics[width=0.45\textwidth]{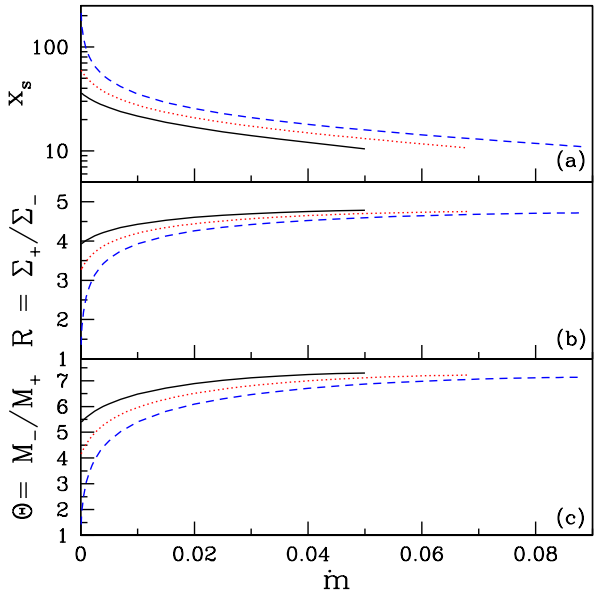}
\end{center}
\caption{
Variation of (a) shock location $x_s$, (b) compression ratio $R$, and (c) shock
strength $\Theta$ as function of $\dot{m}$ for flows injected from
$x_{\rm edge} = 500$ with $\alpha_B = 0.02$, $\beta_{\rm edge} = 1400$ and
${\cal E}_{\rm edge}=1.025 \times 10^{-3}$. Solid, dotted and 
dashed curves represent the results corresponding to 
$\lambda_{\rm edge}= 0.13757\lambda_K$,
$0.13865\lambda_K$ and $0.13972\lambda_K$, respectively.
See text for details. 
}
\end{figure}

Fig. 5, illustrates the various shock properties as function of accretion rate ($\dot{m}$)
for flows injected from a fixed outer edge as $x_{\rm edge} = 500$ with 
$\beta_{\rm edge} = 1400$, energy ${\cal E}_{\rm edge}=1.025\times 10^{-3}$ and
viscosity $\alpha_B = 0.02$. In the upper panel (Fig. 5a), the variation of shock location 
is shown for three different values of angular momentum ($\lambda_{\rm edge}$) at
$x_{\rm edge}$. 
The solid, dotted and dashed curves correspond to flows injected with angular momentum 
$\lambda_{\rm edge}= 0.13757\lambda_K$, $0.13865\lambda_K$ and $0.13972\lambda_K$, 
respectively. From the figure, 
it is evident that a wide range of $\dot{m}$ provides shock induced global accretion solutions. 
For a given $\lambda_{\rm edge}$, the shock position advances towards the horizon
with the increase of accretion rate ($\dot{m}$). The increase of accretion rate enhances 
the efficiency of radiative cooling and the flow loses energy during accretion. Loss of
energy leads to drop in post-shock thermal pressure and hence the shock front moves closer to 
the horizon in order to maintain pressure balance across the shock. When accretion rate
is crossed its critical value ($\dot{m}^{\rm cri}$), standing shock fails to form as the
shock conditions are no longer satisfied. This clearly provides an indication that the
possibility of shock formation reduces with the increase of $\dot{m}$. It is to be noted
that $\dot{m}^{\rm cri}$ does not possess a global value, instead it largely depends on
the accretion flow parameters. Interestingly, when $\dot{m} > \dot{m}^{\rm cri}$,
flow may contain oscillatory shocks, however, the analysis of non-steady shock properties are 
beyond the scope of the present work. In addition, for a given $\dot{m}$, shock front moves
outwards from the horizon when $\lambda_{\rm edge}$ is increased.
This happens due to
the fact that the centrifugal barrier becomes stronger with the increase of
$\lambda_{\rm edge}$ which clearly indicates that the shocks are centrifugally driven.

It is already pointed out that the density and temperature of the PSC is increased 
due to the effect of shock compression. Moreover, the spectral properties of an
accretion disc are directly dependent on the density and temperature distribution
of the flow. Therefore, it is worthy to measure the amount of density and temperature
enhancement across the shock transition. Towards this, we first compute the compression
ratio which is defined as the ratio of the vertically averaged post-shock density
to the pre-shock density ($R=\Sigma_{+}/\Sigma_{-}$) and plot it as function of $\dot{m}$
in Fig. 5b. The flow parameters are chosen as in Fig. 5a. For fixed $\lambda_{\rm edge}$,
$R$ is found to increase monotonically with the increase of $\dot{m}$. This happens because
shock front is pushed towards the horizon with the increase of $\dot{m}$ that boosted the
density compression and subsequently increases the compression ratio. On the contrary, 
for fixed $\dot{m}$, when $\lambda_{\rm edge}$ is increased, shock recedes away due to
the stronger centrifugal barrier causing the decrease of post-shock compression. Since
shock ceases to exist for $\dot{m} > \dot{m}^{\rm cri}$, we observe a cut-off in $R$ for
all the cases. Next, we calculate the strength of the shock ($\Theta$) which is
defined as the ratio of pre-shock Mach number ($M_{-}$) to the post-shock Mach
number ($M_{+}$) and it is essentially measures the temperature jump across the
shock. In Fig. 5c, we show the variation of $\Theta$ as function of $\dot{m}$ for the 
same set of flow parameters as in Fig. 5a. We find that the response of $\Theta$
on the increase of $\dot{m}$ is similar to $R$ as described in Fig. 5b.

\begin{figure}
\begin{center}
\includegraphics[width=0.45\textwidth]{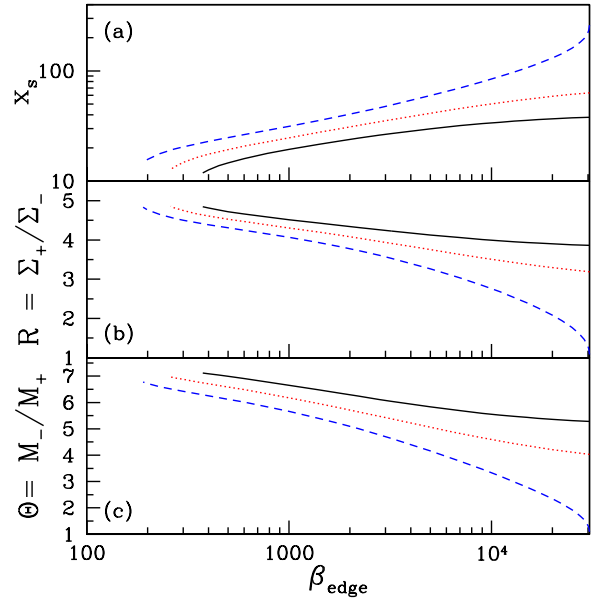}
\end{center}
\caption{
Variation of (a) the shock location $x_s$ (b) shock compression ratio $R$ and (c) shock
strength $\Theta$ as function of $\beta_{\rm edge}$ for different values of
$\lambda_{\rm edge}$. Flows are injected from $x_{\rm edge} = 500$ with
${\cal E}_{\rm edge}=1.025\times 10^{-3}$, $\alpha_B = 0.02$ and
$\dot{m}=0.01$, respectively.  Results depicted with solid, dotted and
dashed curves are for $\lambda_{\rm edge} = 0.13757\lambda_K$,
$0.13865\lambda_K$ and $0.13972\lambda_K$, respectively. See text for details.
}
\end{figure}

In Fig. 6, we proceed to explore the shock properties in terms of $\beta_{\rm edge}$ 
for flows with same outer boundary values, namely, $x_{\rm edge} = 500$, 
${\cal E}_{\rm edge}=1.025\times 10^{-3}$, $\alpha_B = 0.02$ and $\dot{m}=0.01$.
The solid, dotted and dashed curves represent results corresponding to
$\lambda_{\rm edge}= 0.13757\lambda_K$, $0.13865\lambda_K$ and $0.13972\lambda_K$,
respectively. Here also
we observe that the shock front proceeds towards the horizon with the decrease of
$\beta_{\rm edge}$ for all cases having different $\lambda_{\rm edge}$. When 
$\beta_{\rm edge}$ is reduced, the effect of synchrotron cooling is increased 
due to the increase of magnetic activity and therefore, shock moves inward.
However, the indefinite reduction of $\beta_{\rm edge}$ is not possible keeping the 
remaining flow parameters unchanged because under a critical limit 
($\beta_{\rm edge}^{\rm cri}$), shock ceases to exist. As in Fig. 5b-c, here
also we study the variation of compression ratio ($R$) and the shock strength
($\Theta$) as function of $\beta_{\rm edge}$. We find that both $R$ and $\Theta$
display anti-correlation relation with $\beta_{\rm edge}$.

\begin{figure}
\begin{center}
\includegraphics[width=0.45\textwidth]{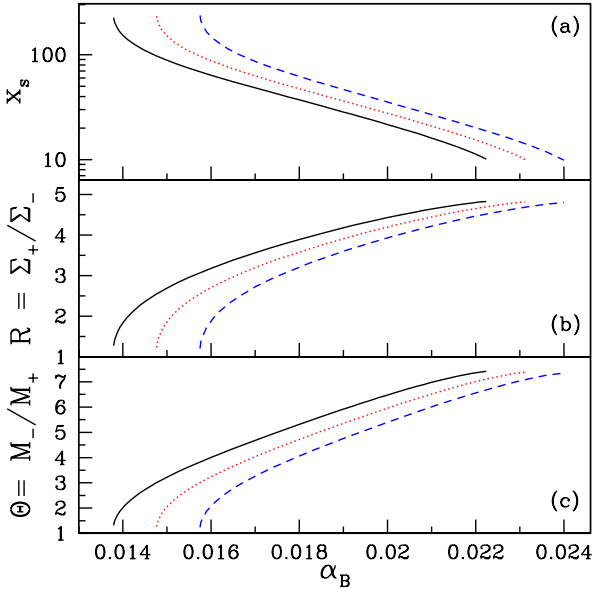}
\end{center}
\caption{
Variation of (a) the shock location $x_s$ (b) shock compression ratio $R$ and (c) shock
strength $\Theta$ as function of viscosity $\alpha_{\rm B}$ for different values of
$\lambda_{\rm edge}$. Flows are injected from $x_{\rm edge} = 500$ with
${\cal E}_{\rm edge}=1.025\times 10^{-3}$, $\beta_{\rm edge} = 1400$ and
$\dot{m}=0.01$. Results drawn with solid, dotted and dashed curves are for
$\lambda_{\rm edge} = 0.13757\lambda_K$, $0.13865\lambda_K$ and $0.13972\lambda_K$, respectively.
See text for details. 
}
\end{figure}

Next, we study the properties of shock wave in terms of viscosity ($\alpha_B$). While doing
so, as before, we choose the flow injection parameter as $x_{\rm edge} = 500$, 
${\cal E}_{\rm edge}=1.025\times 10^{-3}$, $\beta_{\rm edge} = 1400$ and
$\dot{m}=0.01$. We depict the results in Fig. 7 where, solid, dotted and dashed
curve are obtained for $\lambda_{\rm edge} = 0.13757\lambda_K$, $0.13865\lambda_K$ and 
$0.13972\lambda_K$, respectively.
We observe that shock forms for a wide range of $\alpha_B$.
The shock position is reduced with the increase of $\alpha_B$ for all cases having
different $\lambda_{\rm edge}$.  Increase of $\alpha_B$ enhances the angular 
momentum transport outwards that causes the weakening of the centrifugal
barrier and hence the shock is pushed towards the horizon. When $\alpha_B$
is chosen beyond its critical value ($\alpha_B^{\rm cri}$), shock conditions 
are not satisfied and therefore, standing shock ceases to exist. Needless to 
mention that $\alpha_B^{\rm cri}$ largely depends on the other accretion flow
parameters. When $\alpha > \alpha_B^{\rm cri}$, sub-Keplerian flow deviates
from the Keplerian disc very close to the horizon and flow enters into the
black hole after passing through a single sonic point. Further, we calculate
the compression ratio ($R$) and the shock strength ($\Theta$) for flows 
described in Fig. 7a and present the obtained results in Fig. 7b-c. We observe
that both $R$ and $\Theta$ are gradually increased with $\alpha_B$.

\begin{figure}
\begin{center}
\includegraphics[width=0.45\textwidth]{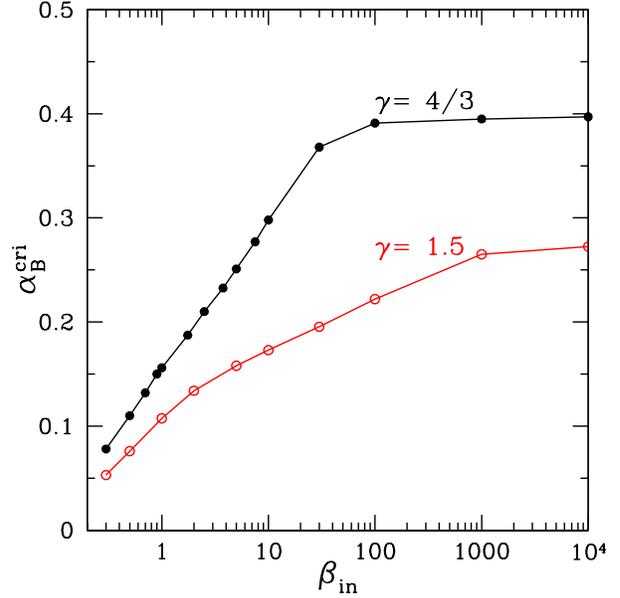}
\end{center}
\caption{
Variation of critical viscosity parameter ($\alpha_B^{cri}$) as function of plasma
$\beta$ at inner sonic point ($\beta_{\rm in}$) that allows standing shocks. Here, we choose 
$\dot{m} = 0.01$. Adiabatic  indices  are marked. See text for details.}
\end{figure}

So far, we have examined the properties of shock induced global accretion solutions for
flows having adiabatic index $\gamma = 4/3$. In reality, the limiting values of $\gamma$
lie in the range between $4/3$ to $5/3$, depending on the ratio of the thermal energy
and the rest energy of the flow \citep{fj02}. Keeping this in mind, we now intend to compute
the critical viscosity parameter ($\alpha^{\rm cri}_{B}$) as function of $\beta_{\rm in}$ that 
allows shocked accretion solution. While doing this, we consider thermally ultra-relativistic
flow ($\gamma \sim 4/3$) and thermally semi-non-relativistic flow ($\gamma \sim 1.5$)
\citep{kc14,YuanNarayan14} and obtain $\alpha^{\rm cri}_{B}$ for both the extreme cases
as depicted in Fig. 8. Here, we assume ${\dot m} = 0.01$.  Filled and open circles joined using
the solid line represent the results corresponding to $\gamma = 4/3$ and $1.5$, respectively.
In a magnetized accretion flow with a given $\alpha_B$, the transport of angular momentum
towards the outer edge of the disc is increased with the decrease of $\beta_{\rm in}$ as
the magnetic pressure contributes to the total pressure. Since shocks under
consideration are centrifugally driven (see Fig. 5), therefore, when $\beta_{\rm in}$ is small,
a lower value of $\alpha_B$ is sufficient to transport the required angular momentum for
shock formation. Evidently, $\alpha^{\rm cri}_B$ possesses lower value for magnetized 
flow. As $\beta_{\rm in}$ is increased, $\alpha^{\rm cri}_B$ is also increased and eventually
approached towards a saturation value corresponding to the gas pressure dominated
flow. For $\gamma = 4/3$,  critical viscosity parameter tends to the asymptotic value 
$\alpha_B^{\rm cri} \sim 0.4$ \citep{Chakrabarti-Das04} and for $\gamma = 1.5$, 
the saturation value is found to be $\alpha_B^{\rm cri} \sim 0.27$ \citep{Das-etal09,Sarkar-Das16}.
We observe that $\alpha_B^{\rm cri}$ is reduced as the flow changes its character from
thermally ultra-relativistic ($\gamma = 4/3$) to thermally semi-non-relativistic limit 
($\gamma = 1.5$). This apparently indicates that the possibility of shock transition seems to be 
feeble when the flow  approaches to the non-relativistic regime. This happens due to the
fact that when $\gamma$ tends to $5/3$, flow possesses only single sonic point
\citep{Chakrabarti-Das04},  thereby reducing the possibility of shock formation.
 
\begin{figure}
\begin{center}
\includegraphics[width=0.45\textwidth]{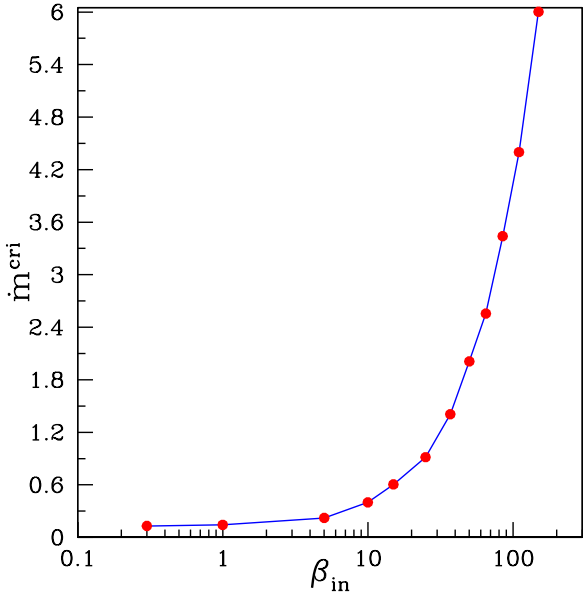}
\end{center}
\caption{
Variation of critical accretion rate $\dot{m}^{\rm cri}$ that allows standing shock as function of 
$\beta_{\rm in}$. Here,  viscosity parameter is chosen as $\alpha_B = 0.01$. See text for details.
}
\end{figure}

We continue our investigation to calculate the critical accretion rate of the flow that provides 
the global accretion solutions including standing shock. In Fig. 9, we present the variation of
critical accretion rate ($\dot{m}^{\rm cri}$) as function of plasma $\beta$ measured 
at the inner sonic point ($\beta_{\rm in}$). Here, we choose viscosity parameter
as $\alpha_B = 0.01$. In this analysis, we have chosen the wide range of $\beta$ where, 
$\beta < 1$ represents the magnetic pressure dominated flow and $\beta > 1$ denotes the
gas pressure dominated flow. Moreover, we consider
synchrotron cooling as the effective radiative mechanism active in the flow. Since synchrotron
process depends on both the strength of the magnetic fields and the density of accreting matter,
one can obtain the required cooling effect by suitably tuning the density and the magnetic fields
together.  When $\beta_{\rm in} < 1$, the inner part of the disc becomes magnetically dominated
as the disc is threaded with strong magnetic fields as compared to the gas dominated disc
and therefore, significant cooling effect can be achieved even with small accretion rate. Thus, 
for magnetically dominated flow, we obtain small values of $\dot{m}^{\rm cri}$. 
For example, we obtain the magnetic field at $x_{\rm in}=2.5234$ as 
$B (x_{\rm in})=1.453 \times 10^{7}~{\rm Gauss}$ for 
$\lambda_{\rm in} = 0.978185\lambda_K$, $\alpha_{B}=0.01$,
$\beta_{\rm in}= 0.3$, $\dot{m}^{\rm cri}=0.13$ and $M_{BH}=10M_{\odot}$, respectively. As we
gradually increase the value of $\beta_{\rm in}$, the effect of magnetic fields becomes
weaker that eventually decreases the cooling effect in the flow. Hence, flow can sustain
standing shock even with relatively large accretion rate. When $\beta \gg 1$, magnetic
fields are very weak leading to the negligible cooling effect in the flow. Therefore,
we are effectively left  with a flow where $\dot{m}^{\rm cri}$ tends to become
independent of $\beta_{\rm in}$ as observed in the figure.

\begin{figure}
\begin{center}
\includegraphics[width=0.45\textwidth]{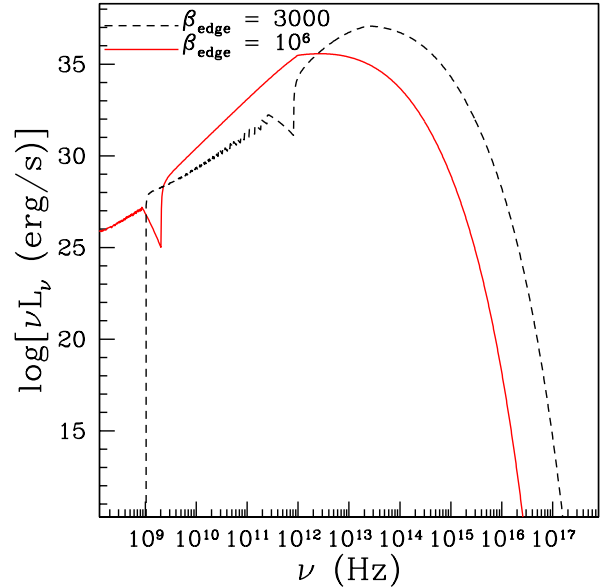}
\end{center}
\caption{Typical spectra from an accretion disc around a black hole of mass
$M = 10 M_{\odot}$ for a strong and weak accretion shock located at $x_s =
13.14$ and $72.12$ respectively.
The two different shock locations correspond to flows injected from the
outer edge with two different values
of $\beta$ as indicated in the figure. See text for details.
}
\end{figure}

In Fig. 10, we present the typical spectrum for shocked accretion flows injected from
outer edge $x_{\rm edge} = 500$ with angular momentum 
$\lambda_{\rm edge}=0.13865\lambda_K$ and energy 
${\cal E}_{\rm edge}=1.025 \times 10^{-3}$. Here, we consider
viscosity as $\alpha_B = 0.02$ and accretion rate as ${\dot m}=0.1$. For representation, 
black hole mass is chosen as $M_{BH} = 10 M_{\odot}$. 
Considering $\beta_{\rm edge} = 10^6$, we obtain the global accretion solution 
where standing shock is formed at $x_s = 72.12$. Following the works of 
\citet{MandalChakrabarti05,ChakrabartiMandal06}, we compute the disc synchrotron 
spectrum corresponding to this accretion
solution and present it in Fig. 10 using solid curve. 
Further, we increase the effect of magnetic fields by setting
$\beta_{\rm edge} = 3000$ and inject the flow keeping the remaing flow variables
unaltered.
We observe that the flow encounters standing shock transition at $x_s = 13.14$.
We then compute the disc spectrum as before and depict it in Fig. 10 using 
dashed curve. In both the cases, the pre-shock (low energy radiation) and
post shock (high energy radiation) synchrotron contributions are well separated across the
sharp discontinuity due to sudden jump in temperature, density and
magnetic field at the shock. The spikes in pre-shock contribution represent the
cumulated cyclotron lines coming from different disc annuli. Evidently, lower
$\beta$ at outer boundary (dashed curve) corresponds to higher magnetic fields
and therefore, the flow will be radiatively more efficient. Accordingly, the spectrum
is shifted towards high energy for large magnetic field value. This essentially indicates
that the disc
is making a transition to a brighter hard state with the decrease of $\beta_{\rm edge}$. 
The above findings is in agreement with the results of \citet{oda12}, where
the brightening of the hard state is reported for disc that makes transition to
the low-$\beta$ state.

\begin{figure}
\begin{center}
\includegraphics[width=0.45\textwidth]{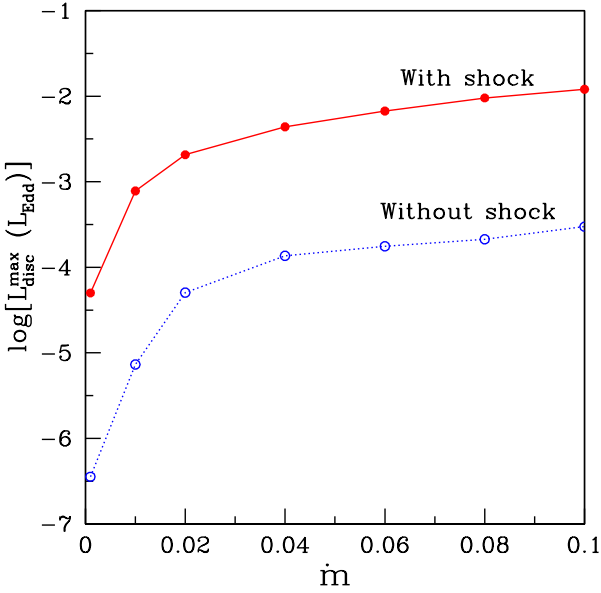}
\end{center}
\caption{
Variation of maximum disc luminosity $L_{disc}^{max}$ as a 
function of accretion rate $\dot{m}$. Filled circles connected with solid line denote the
results for shocked accretion flows whereas the open circles joined with dotted line
are for shock free accretion solutions. See text for details.}
\end{figure}

In the course of our investigation of accretion flow properties, we next put an effort
to calculate the disc luminosity. Since we use the synchrotron emission as the potential
cooling mechanism while modeling the accretion flows around black holes, the total
surface disc 
luminosity ($L_{\rm disc}$) is estimated as,

$$
L_{\rm disc} = 4\pi \int_{x_i}^{x_{\rm f}} Q^{-}x dx
$$
where, the limit $x_{i}$ refers to the location just outside the horizon and $x_{f}$ stands for
the outer edge of the disc, respectively, and $Q^{-}$ denotes the synchrotron cooling rate.
Here, for a given accretion rate ($\dot{m}$), we  compute the
maximum disc luminosity ($L_{\rm disc}^{\rm max}$) 
for shock and shock free accretion solutions by freely varying the remaining flow parameters
and show its variation in units of Eddington
luminosity as function of $\dot{m}$ in Fig. 11. Filled circles joined with
the solid line represent the results corresponding to the global accretion solutions
including shocks and the open circles connected with the dotted curve are for shock
free accretion solutions. Overall, we observe that $L_{\rm disc}^{\rm max}$ increases
with $\dot{m}$ for all cases. This happens, simply because the increase of $\dot{m}$
manifests the rise of the flow density that eventually allows the flow to cool down
more efficiently. Moreover, we find that for a fixed $\dot{m}$, $L_{\rm disc}^{\rm max}$
remains always higher for shocked accretion flows compared to the shock free flows.
In other words, according to our model, the accretion flow containing shock waves are
radiatively more efficient than the flows having no shock. Therefore, it is fairly indicative
that the shocked accretion solutions seems to be potentially more preferred over the
shock free solutions in explaining the energetics of black hole sources. 

\begin{figure}
\begin{center}
\includegraphics[width=0.45\textwidth]{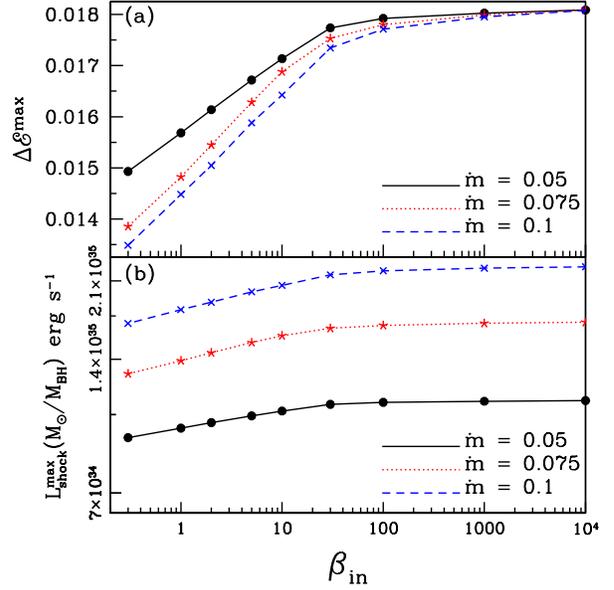}
\end{center}
\caption{\label{lab:delE_max_vs_bta_in_data} Variation of maximum energy dissipation at shock 
($\Delta \mathcal{E}^{\rm max}$) as a function of $\beta$ at inner sonic point ($\beta_{\rm in}$) 
for three different values of accretion rates, $\dot{m} = 0.05$, $0.075$ and $0.1$ denoted by 
solid, dotted and dashed curves, respectively.}
\end{figure}

Until now, we have studied the various properties of accretion shock waves around black
holes. These shocks are non-dissipative in nature as the specific energy remains conserved
across the shock front \citep{c89}. Indeed, shocks of this kind are radiatively inefficient
as well. Apparently, the realistic shock waves are likely to be dissipative, where a part of
accreting energy is escaped from the shock location through the disc surfaces.
This essentially causes the
reduction of the overall specific energy profile in the PSC region \citep{Singh-Chakrabarti11}.
The plausible mechanism that perhaps regulates the energy dissipation at the shock  is
the thermal Comptonization process \citep{ct95} that eventually reduce the thermal energy 
in the PSC. Based on the above consideration, we assume that the loss of energy across
the shock front is proportional to the temperature difference between the immediate pre-shock
and post shock flow and therefore, the energy loss across the shock front is calculated as
\citep{Das-etal10},
$$ 
\Delta \mathcal{E} = f n (a_+^2 - a_-^2),
\eqno(25)
$$
where, $a_{-}$ and $a_{+}$ denote the immediate pre-shock and post-shock sound speed and
$f$ represents the fraction of the available thermal energy lost at shock. In this analysis, we
treat $f$ as a free parameter and chose its value as $f=0.998$ for representation. 
Evidently, this is the measure of energy dissipation across the shock \citep{Das-etal10,
Singh-Chakrabarti11,sarkar2013,kc13}. We then compute the maximum energy dissipation
($\Delta \mathcal{E}^{\rm max}$) by freely varying all the flow parameters and plot it
as function of $\beta_{\rm in}$ for various values of accretion rates in Fig. 12a. Here,
solid, dotted and dashed curves represent the results corresponding to 
$\dot{m} = 0.05$, $0.075$ and $0.1$, respectively. We observe that for a given
accretion rate (${\dot m}$), $\Delta \mathcal{E}^{\rm max}$ initially shows gradual
increment with $\beta_{\rm in}$ which eventually approaches towards a saturation
value in the gas pressure dominated domain. Meanwhile, it was shown by numerical
simulation \citep{machida06} that in the magnetically dominated state, accretion disc
becomes cool as compared to a gas pressure dominated disc. This clearly indicates
that the gas pressure dominated flow is expected to possess higher thermal energy
content and accordingly, we obtain enhanced $\Delta \mathcal{E}^{\rm max}$ when
$\beta_{\rm in}$ is larger. Moreover, as ${\dot m}$ is increased, cooling becomes more
effective in the flow resulting the lowering of thermal energy content. Therefore, for a given
$\beta_{\rm in}$, $\Delta \mathcal{E}^{\rm max}$ remains smaller for flows having
higher ${\dot m}$. When $\beta_{\rm in}$ is large, the effect of synchrotron cooling
due to the increase of accretion rate becomes practically insignificant. Hence,  
when $\beta_{\rm in}\rightarrow 10^{4}$, $\Delta \mathcal{E}^{\rm max}$ asymptotes 
to a saturation value irrespective to the flow accretion rate.

In the next, we intend to infer the utility of dissipated energy at the shock
($\Delta \mathcal{E}$) in terms of the observable quantities. Indeed, 
a portion of the total usable energy available at the PSC is the same dissipative energy 
$\Delta \mathcal{E}$. Since jets are likely to be launched  from the PSC, a part of
this energy is used in the process of  jet generation.
Keeping this in mind, we calculate the maximum kinetic power lost by the disc corresponding to 
$\Delta \mathcal{E}^{\rm max}$ in terms of shock luminosity as $L^{\rm max}_{\rm shock} =
{\dot M} \times \Delta \mathcal{E}^{\rm max} \times c^2 {~\rm erg~s^{-1}}$
\citep{Le-Becker04,Le-Becker05,Sarkar-Das16}. Here, ${\dot M} (\equiv {\dot m}{\dot M}_{edd})$ represents the
accretion rate in physical units. With this, in Fig. 12b, we show the variation of
the maximum shock luminosity (scaled with black hole mass) as a function of $\beta_{\rm in}$
for various accretion rates ($\dot{m}$). As before, solid, dotted and dashed curves represent the
results corresponding to $\dot{m} = 0.05$, $0.075$ and $0.1$, respectively. Since we
calculate the maximum shock luminosity using $\Delta \mathcal{E}^{\rm max}$,
the obtained results apparently depend on the accretion rate although it manifest
the variations similar to $\Delta \mathcal{E}^{\rm max}$ (see Fig. 12a).
Consequently, we observe that for a given $\beta_{\rm in}$,  the maximum
shock luminosity is increased with $\dot{m}$.
Based on the above findings, we point out that our model calculation of shock
luminosity  (as depicted in Fig. 12b) can be readily used to understand the
observational findings of core radio luminosity values associated with the
black hole sources.

\section{Conclusions}

In this paper, we study the effect of synchrotron cooling in a  magnetized advective
accretion flow around a non-rotating black hole. 
While investigating the various properties of the accretion flow, since the origin of viscosity
and the exact mode of angular momentum transport in accretion discs is still remain 
inconclusive, we rely on the numerical simulation results of  \citet{machida06} and assume
that the $x\phi$-component of the Maxwell stress is proportional to the total pressure
of the accreting matter. During accretion, flow changes its sonic state from subsonic to
supersonic to become transonic before falling into the black hole. The position where
flow becomes transonic is called as the sonic point. Depending on the input parameters,
flow may pass through the multiple sonic points and the flow of this kind has
the potential to exhibit shock phenomenon. Meanwhile, \citet{oda07,oda12} studied
the global accretion solutions of magnetically supported accretion discs around stationary
black holes. In these works, authors considered the accretion solutions that pass
through the inner sonic point only.  Essentially, these solutions are the subset of the
generalized transonic accretion solutions as they ignored the flows containing multiple 
sonic points.
Very recently, \citet{Sarkar-Das16} studied the properties of the shocked accretion flow
considering bremsstrahlung cooling where magnetic field strength was assumed to be moderate
throughout the flow. In the present work, we calculate the shock induced global accretion 
solution for flows having wide range of plasma $\beta$ parameters as
$0.3 \le \beta_{\rm in} \le \infty$. With this, we further examine the effects of the dissipation 
parameters, namely viscosity ($\alpha_B$) and accretion rate ($\dot{m}$),
on the properties of global accretion  solutions that contains shock waves. Such a study is
important in the sense that the dissipation processes are likely to influence the 
spectral and timing properties of the radiation emitted from the disc
\citep{Chakrabarti-Manickam00,Nandi-etal01a,Nandi-etal01b,
Nandi-etal12,Radhika-Nandi14,Iyer-etal15}.

Our main concern here is to 
obtain the global magnetized accretion solution in presence of synchrotron cooling that 
contains shock wave (Fig. 2). For an accretion flow injected from a fixed outer edge, the dynamics of
 the shock front is regulated by cooling parameters, namely accretion rate (${\dot m}$) and
plasma $\beta$. Due to
the presence of discontinuous shock transition, post-shock flow (\ie PSC) is compressed resulting
a hot and dense PSC. Therefore, when accretion rate and/or magnetic field are increased, cooling
efficiency is enhanced causing the reduction of thermal pressure at PSC. This eventually compels the
shock front to move towards smaller distance in order to balance the total pressure across the shock (Fig. 3).  In 
our model solution, we find that the accretion flow is very hot, magnetized and optically as well
as geometrically thin in the inner part of the disc (Fig. 4). This eventually allows the hard radiation
to escape from PSC with ease. Moreover, we find that shock induced global accretion solutions
are not the discrete solutions as shock forms for a wide range of flow parameters.
 Interestingly, above a critical limit of cooling parameters 
(${\dot m}^{\rm cri}$ and $\beta^{\rm cri}_{\rm edge}$), PSC ceases to exist as the standing
shock conditions fails to satisfy in presence of excess cooling (Fig. 5-6).

Next, we put an effort to examine the properties of global shock solutions in both gas pressure
as well as magnetic pressure dominated flow. While doing this, we calculate the critical viscosity
parameter ($\alpha_B^{\rm cri}$) that caters standing shock waves. Here, we consider two different
values of adiabatic index that represents accretion flow lying in the range between ultra-relativistic 
($\gamma = 4/3$) to semi-non-relativistic ($\gamma = 1.5$) domain. We observe that in all cases,
$\alpha_B^{\rm cri}$ initially increases with $\beta_{\rm in}$ and asymptotically approaches to the
saturation values $\sim 0.4$ (for $\gamma = 4/3$) and  $\sim 0.27$ (for $\gamma = 1.5$), respectively
for gas pressure dominated flow  \citet{Chakrabarti-Das04,king2007,Das-etal09,kc13}. Our steady
model eventually establishes the fact that global shocks accretion solution can be obtained for
fairly high viscosity parameter (Fig. 8). 

We have further estimated the critical accretion rate (${\dot m}^{\rm cri}$) that provides the
global accretion solution including shock waves. We find that ${\dot m}^{\rm cri}$ is small
for magnetically dominated flow as it is adequate to provide the required cooling efficiency
that can sustain the standing shock in the accretion flow. As the strength of the magnetic
field decreases, ${\dot m}^{\rm cri}$ gradually increases and ultimately ${\dot m}^{\rm cri}$
tends to become independent when magnetic fields are very weak leading to the gas pressure
dominated flow (Fig. 9).

In Fig. 10, we have explored the typical spectrum of a magnetized accretion disc around
black holes. For the purpose of representation, we consider two accretion flows having
different magnetic field strengths. We find that the spectrum moves towards the high energy
when magnetic field is large. This clearly indicates that the disc makes transition to the
the brighter hard state as the disc becomes magnetically dominated. In addition, we 
compare the disc luminosities for accretion solutions with and without shock and observe
that shock accretion solutions are radiatively more efficient than the shock free solutions.
This provides a possible hint that global shock solutions are potentially preferred over 
the shock free solutions in understanding the energetics of the black hole sources.

Another important findings we examine in our steady state model when shocks under
consideration are assumed to be dissipative in nature. In this circumstances, a part of
the accreting energy is liberated at the shock which is allowed to escape through the
disc surface. Interestingly, this available energy dissipated at shock can be used in
powering the jets \citep{Le-Becker04,Le-Becker05,Das-etal09}. Following this, we
self-consistently calculate the maximum radiative luminosity at shock 
($L^{\rm max}_{\rm shock}$)
corresponding to $\Delta \mathcal{E}^{\rm max}$ as an observable quantity
and argue that the obtained results may be used to explain the observational
findings of radio luminosities corresponding to the black hole sources. 

At the end, we would like to point out that the present steady state formalism is
developed based on several assumptions. For simplicity, we adopt pseudo-Newtonian
potential to describe the space time geometry around the black hole. We consider
the adiabatic constant of the flow as global constant although it should be estimated
self-consistently from the thermal properties of the accreting matter. In general,
since the inner part of the disc is very hot and the radiative cooling time of
relativistic electrons are shorter than the non-relativistic ions, the accreting
plasma is expected to be characterized by two temperature flow. As the 
implementation of the above issues are beyond the scope of the present paper,
all these aspects will be the subject of our future study and will be reported
elsewhere.

\section*{Acknowledgements}
Authors thank the anonymous referee for useful comments and constructive suggestions.





\begin{thebibliography}{99}

\bibitem[\protect\citeauthoryear{Akizuki \& Fukue}{2006}]{Akizuki-Fukue06}
Akizuki C., Fukue J., 2006, PASJ, 58, 469

\bibitem[\protect\citeauthoryear{Aktar \etal}{2015}]{Aktar-etal15}
Aktar R., Das S., Nandi A., 2015, \mnras, 453, 3414

\bibitem[\protect\citeauthoryear{Balbus \& Hawley}{1991}]{Balbus-Hawley91}
Balbus S., Hawley J. F., 1991, \apj, 376, 214

\bibitem[\protect\citeauthoryear{Balbus \& Hawley}{1998}]{Balbus-Hawley98}
Balbus S.~A., Hawley J.~F., 1998, RvMP, 70, 1

\bibitem[\protect\citeauthoryear{Becker \& Kazanas}{2001}]{Becker-Kazanas01}
Becker P. A., Kazanas D., 2001, \apj, 546, 429

\bibitem[\protect\citeauthoryear{Bisnovatyi-Kogan \& Ruzmaikin}{1974}]{Bisnovatyi-Kogan-Ruzmaikin74} 
Bisnovatyi-Kogan G.~S., Ruzmaikin A.~A., 1974, Ap\&SS, 28, 45 

\bibitem[\protect\citeauthoryear{Bisnovatyi-Kogan \& Blinnikov}{1976}]{Bisnovatyi-Kogan-Blinnikov76}
Bisnovatyi-Kogan G. S., Blinnikov S. I., 1976, Sov. Astron. Lett., 2, 191

\bibitem[\protect\citeauthoryear{Begelman \& Pringle}{2007}]{Begelman-Pringle07}
Begelman M. C., Pringle J.E., 2007, \mnras, 375, 1070

\bibitem[\protect\citeauthoryear{Bu \etal}{2009}]{bu09}
Bu D. F., Yuan F., Xie F. G., 2009, \mnras, 392, 325

\bibitem[\protect\citeauthoryear{Chakrabarti}{1989}]{c89}
Chakrabarti S. K., 1989, \apj, 347, 365

\bibitem[\protect\citeauthoryear{Chakrabarti \& Titarchuk}{1995}]{ct95}
Chakrabarti S K., Titarchuk L., 1995, \apj, 455, 623.

\bibitem[\protect\citeauthoryear{Chakrabarti}{1996}]{Chakrabarti96}
Chakrabarti S. K., 1996, \apj, 464, 664

\bibitem[\protect\citeauthoryear{Chakrabarti}{1999}]{c99}
Chakrabarti S. K., 1999, \aap, 351, 185

\bibitem[\protect\citeauthoryear{Chakrabarti \& Das}{2004}]{Chakrabarti-Das04}
Chakrabarti S. K., Das S., 2004, \mnras, 349, 649

\bibitem[\protect\citeauthoryear{Chakrabarti \& Manickam}{2000}]{Chakrabarti-Manickam00}
Chakrabarti S. K., Manickam S. G., 2000, \apj, 531, L41

\bibitem[\protect\citeauthoryear{Chakrabarti \& Mandal}{2006}]{ChakrabartiMandal06}
Chakrabarti S.~K., Mandal S., 2006, ApJ, 642, L49

\bibitem[\protect\citeauthoryear{Chattopadhyay \& Chakrabarti}{2000}]{cc00}
Chattopadhyay I., Chakrabarti S. K., 2000, IJMPD, 9, 717

\bibitem[\protect\citeauthoryear{Chattopadhyay \& Chakrabarti}{2002}]{cc02}
Chattopadhyay I., Chakrabarti S. K., 2002, \mnras, 333, 454

\bibitem[\protect\citeauthoryear{Chattopadhyay \& Das}{2007}]{cd07}
Chattopadhyay I., Das S., 2007, \na, 12, 454

\bibitem[\protect\citeauthoryear{Chattopadhyay \& Chakrabarti}{2011}]{cc11}
Chattopadhyay I., Chakrabarti S. K., 2011, IJMPD, 20, 1597

\bibitem[\protect\citeauthoryear{Das \etal}{2001a}]{dcc01}
Das S., Chattopadhyay, I. \& Chakrabarti, S. K., 2001a, \apj, 557, 983

\bibitem[\protect\citeauthoryear{Das \etal}{2001b}]{dcnc01}
Das S., Chattopadhyay I., Nandi A., Chakrabarti S. K., 2001b, A\&A 379, 683

\bibitem[\protect\citeauthoryear{Das}{2007}]{d07}
Das S., 2007, MNRAS, 376, 1659

\bibitem[\protect\citeauthoryear{Das \etal}{2009}]{Das-etal09}
Das S., Becker P. A., Le T., 2009, \apj, 702, 649

\bibitem[\protect\citeauthoryear{Das \etal}{2010}]{Das-etal10}
Das S., Chakrabarti S. K., Mondal S., 2010, \mnras, 401, 2053

\bibitem[\protect\citeauthoryear{Das \etal}{2014a}]{Das-etal14}
Das S., \etal, 2014a, \mnras, 442, 251

\bibitem[\protect\citeauthoryear{Das \etal}{2014b}]{dcns14}
Das S., Chattopadhyay I., Nandi A., Sarkar B., 2014b, BASI 42, 39

\bibitem[\protect\citeauthoryear{Frank \etal}{2002}]{fj02}
Frank J., King A., Raine D. J., 2002, `Accretion Power in Astrophysics', Cambridge, UK: Cambridge University Press.

\bibitem[\protect\citeauthoryear{Fukue}{1990}]{fukue90}
Fukue J., 1990, \pasj, 42, 793

\bibitem[\protect\citeauthoryear{Hirose \etal}{2006}]{hirose2006}
Hirose S., Krolik J.~H., Stone J.~M., 2006, ApJ, 640, 901

\bibitem[\protect\citeauthoryear{Iyer, Nandi, \& Mandal}{2015}]{Iyer-etal15}
Iyer N., Nandi A., Mandal S., 2015, ApJ, 807, 108

\bibitem[\protect\citeauthoryear{Johansen \& Levin}{2008}]{johansen2008}
Johansen, A., Levin, Y., 2008, A\&A, 490, 501

\bibitem[\protect\citeauthoryear{Khanna \& Camenzind}{1996}]{Khanna-Camenzind96}
Khanna R., Camenzind M., 1996, A\&A, 307, 665

\bibitem[\protect\citeauthoryear{King \etal}{2007}]{king2007}
King A. R., Pringle J. E. \& Livio N., 2007, \mnras, 376, 1740

\bibitem[\protect\citeauthoryear{Krolik \etal}{2007}]{krolik2007}
Krolik J. H., Hirose S. \& Blaes O., 2007, \apj, 664, 1045

\bibitem[\protect\citeauthoryear{Kumar \& Chattopadhyay}{2013}]{kc13}
Kumar R., Chattopadhyay I., 2013, \mnras, 430, 386

\bibitem[\protect\citeauthoryear{Kumar \& Chattopadhyay}{2014}]{kc14}
Kumar R., Chattopadhyay I., 2014, \mnras, 443, 3444

\bibitem[\protect\citeauthoryear{Le \& Becker}{2004}]{Le-Becker04}
Le T., Becker P. A., 2004, \apjl, 617, 25

\bibitem[\protect\citeauthoryear{Le \& Becker}{2005}]{Le-Becker05}
Le T., Becker P. A., 2005, \apjl, 632, 476

\bibitem[\protect\citeauthoryear{Lynden-Bell}{1969}]{bl69}
Lynden-Bell D., 1969, Nature, 223, 690

\bibitem[\protect\citeauthoryear{Machida \etal}{2006}]{machida06}
Machida M., Nakamura K. E., Matsumoto R., 2006, \pasj, 58, 193.

\bibitem[\protect\citeauthoryear{Mandal \& Chakrabarti}{2005}]{MandalChakrabarti05}
Mandal S., Chakrabarti S.~K., 2005, A\&A, 434, 839

\bibitem[\protect\citeauthoryear{Matsumoto \etal}{1984}]{mkfo84}
Matsumoto R., Kato S., Fukue J., Okazaki A.~T., 1984, PASJ, 36, 71

\bibitem[\protect\citeauthoryear{Miller \etal}{2006}]{m06}
Miller J.~M., Raymond J., Fabian A., Steeghs D., Homan J., Reynolds C., van der Klis M., Wijnands R., 2006, Natur, 441, 953

\bibitem[\protect\citeauthoryear{Mosallanezhad \etal}{2014}]{mosall14}
Mosallanezhad A., Abbassi S., Beiranvand N., 2014, \mnras, 437, 3112

\bibitem[\protect\citeauthoryear{Narayan \etal}{1997}]{nar97}
Narayan R., Kato S., Honma F., 1997, \apj, 476, 49.

\bibitem[\protect\citeauthoryear{Nandi \etal}{2001a}]{Nandi-etal01a} 
Nandi A., Chakrabarti S.~K., Vadawale S.~V., Rao A.~R., 2001a, A\&A, 380, 245

\bibitem[\protect\citeauthoryear{Nandi \etal}{2001b}]{Nandi-etal01b}
Nandi A., Manickam S.~G., Rao A.~R., Chakrabarti S.~K., 2001b, MNRAS, 324, 267

\bibitem[\protect\citeauthoryear{Nandi \etal}{2012}]{Nandi-etal12}
Nandi A., Debnath D., Mandal S., Chakrabarti S.~K., 2012, A\&A, 542, A56

\bibitem[\protect\citeauthoryear{Oda \etal}{2007}]{oda07}
Oda H., Machida M., Nakamura K. E., Matsumoto R., 2007, \pasj, 59, 457

\bibitem[\protect\citeauthoryear{Oda \etal}{2010}]{oda10}
Oda H., Machida M., Nakamura K. E., Matsumoto, R., 2010, \apj, 712, 639

\bibitem[\protect\citeauthoryear{Oda \etal}{2012}]{oda12}
Oda H., Machida M., Nakamura K. E., Matsumoto R., Narayan, R., 2012, \pasj, 64, 15

\bibitem[\protect\citeauthoryear{Paczy\'nski \& Wiita}{1980}]{pw80}
Paczy\'nski B., Wiita P. J., 1980, \aap, 88, 23.

\bibitem[\protect\citeauthoryear{Radhika \& Nandi}{2014}]{Radhika-Nandi14}
Radhika D., Nandi A., 2014, AdSpR, 54, 1678

\bibitem[\protect\citeauthoryear{Sadowski}{2016}]{sadowski16}
Sadowski A., 2016, MNRAS, 459, 4397

\bibitem[\protect\citeauthoryear{Samadi \etal}{2014}]{sam14}
Samadi M., Abbassi S., Khajavi M., 2014, \mnras, 437, 3124

\bibitem[\protect\citeauthoryear{Sarkar \& Das}{2013}]{sarkar2013}
Sarkar B., Das S., 2013, ASInC, 8, 143

\bibitem[\protect\citeauthoryear{Sarkar \& Das}{2015}]{sarkar2015}
Sarkar B., Das S., 2015, ASInC, 12, 91

\bibitem[\protect\citeauthoryear{Sarkar \& Das}{2016}]{Sarkar-Das16}
Sarkar B., Das S., 2016, MNRAS, 461, 190
 
\bibitem[\protect\citeauthoryear{Shakura \& Sunyaev}{1973}]{ss73}
Shakura N. I., Sunyaev R. A., 1973, \aap, 24, 337.

\bibitem[\protect\citeauthoryear{Shapiro \& Teukolsky}{1983}]{shte83} 
Shapiro S. L., Teukolsky S. A., 1983, Black Holes, White Dwarfs and 
Neutron Stars: The Physics of Compact Objects, A Wiley-Interscience Publication, New York.

\bibitem[\protect\citeauthoryear{Singh \& Chakrabarti}{2011}]{Singh-Chakrabarti11}
Singh C. B., Chakrabarti S. K., 2011, \mnras, 410, 2414

\bibitem[\protect\citeauthoryear{Torkelsson \& Brandenburg}{1994}]{torkel94}
Torkelsson U., Brandenburg A., 1994, \aap, 283, 677.

\bibitem[\protect\citeauthoryear{Yuan \& Narayan}{2014}]{YuanNarayan14}
Yuan F., Narayan R., 2014, ARA\&A, 52, 529            

\end{thebibliography}




%
%


\bsp	
\label{lastpage}
\end{document}